\documentclass[twocolumn,showpacs,aps,groupedaddress,prd,eqsecnum]{revtex4}
\usepackage{amsfonts,latexsym,eucal}
\usepackage{epsfig}

\def\beq{\begin{equation}}
\def\eeq{\end{equation}}
\def\bea{\begin{eqnarray}}
\def\eea{\end{eqnarray}}
\def\nn{\nonumber\\}
\def\pa{\partial}
\def\ra{\rightarrow}
\def\bp{\mbox{\boldmath$\phi$}}

\def\tp{\tilde{\phi}}
\def\ds{\displaystyle}

\begin{document}

\draft
\title{Gauged Q-balls in the Affleck-Dine mechanism}
\author{Takashi Tamaki}
\email{tamaki@ge.ce.nihon-u.ac.jp}
\affiliation{Department of Physics, General Education, College of Engineering, 
Nihon University, Tokusada, Tamura, Koriyama, Fukushima 963-8642, Japan}
\author{Nobuyuki Sakai}
\email{nsakai@yamaguchi-u.ac.jp}
\affiliation{Faculty of Science, Yamaguchi University, Yamaguchi 753-8512, Japan}

\begin{abstract}
We consider gauged Q-balls in the gravity-mediation-type model in the Affleck-Dine mechanism, which is described by the potential $V_{\rm grav.}(\phi):=(m_{\rm grav.}^2/2)\phi^2\left[1+K\ln(\phi/M)^2\right]$ with $K<0$.
In many models of gauged Q-balls, which were studied in the literature, there are upper limits for charge and size of Q-balls due to repulsive Coulomb force.
In the present model, by contrast, our numerical calculation strongly suggests that stable solutions with 
any amount of charge and size exist.
As the electric charge $Q$ increases,  the field configuration of the scalar field becomes shell-like; 
because the charge is concentrated on the surface, the Coulomb force does not destroy the Q-ball configuration.
These properties are analogous to those in the V-shaped model, which was studied by Arod\'z and Lis.
We also find that for each $K$ there is another sequence of unstable solutions, which is separated from the other sequence of the stable solutions.
As $|K|$ increases, the two sequences approach; eventually at some point in $-1.07<K<-1.06$, the ``recombination" of the two sequences takes place.
\end{abstract}

\pacs{03.75.Lm, 11.27.+d}
\maketitle

\section{Introduction}
Among nontopological solitons, Q-balls \cite{Col85} have attracted much attention because
they can exist in all supersymmetric extensions of the Standard Model \cite{Kus97b-98}.
Specifically, they can be produced efficiently in the Affleck-Dine (AD) mechanism \cite{AD} and could be responsible 
for baryon asymmetry \cite{SUSY} and dark matter \cite{SUSY-DM}. 
Q-balls can also influence the fate of neutron stars \cite{Kus98}. 
Based on these motivations, stability of Q-balls has been intensively 
studied \cite{stability,PCS01,SS}.
These studies have also been extended to general relativistic analysis \cite{TS} and to different-shaped solitons \cite{SIN,TS2}.

Another natural extension is introducing gauge coupling into a U(1) scalar field.
Because such a field is equivalent to electromagnetic field, the conserved charge $Q$ becomes electric charge, and therefore the Coulomb repulsion is expected to disturb formation of large Q-balls.
In fact, Lee {\it et al.} \cite{Lee} began to study gauged Q-balls with the potential,
\beq\label{V4}
V_4(\phi):={m^2\over2}\phi^2-\lambda\phi^4+\frac{\phi^6}{M^2} 
~~~{\rm with} ~~~ m^2,~\lambda,~M^2>0, 
\eeq
and showed that there is a maximum charge and size.
To construct large Q-balls, Anagnostopoulos  {\it et al.} \cite{AAFT} introduced fermions with charge of the opposite sign.
Li  {\it et al.} \cite{LHL} assumed a different potential, a piecewise parabolic function, and Deshaies-Jacques and MacKenzie \cite{DM} supposed the Maxwell-Chern-Simons theory with the $V_4$ potential (\ref{V4}) in the 2+1 dimensional spacetime; it was shown that there is a maximum charge and size of Q-balls in both models.

Arod\'z and Lis \cite{Arodz} considered gauged Q-balls with the V-shaped potential,
\beq\label{VV}
V_{\rm V}(\phi):=\lambda\frac{|\phi |}{\sqrt{2}}
~~~{\rm with} ~~~ \lambda >0, 
\eeq
Because its three-dimensional plot has the form of a cone, it would be more appropriate to call it the cone-shaped potential.
In addition to normal Q-balls, which have a maximum charge, they found a new type of solutions, Q-shells.
Q-shell solutions are obtained in such a way that the scalar field and the gauge field are assumed to be constant within a certain sphere $r<r_0$ and the field equations are solved numerically for $r>r_0$. 
Because the electric charge is concentrated on the shell, large Q-balls with any amount of charge can exist without additional fermions. 
Thus this model overcomes the difficulty of the $V_4$ model.
However,  there is another drawback that it is so simplified and singular at $\phi=0$. 

In this paper we address the question whether such large gauged Q-balls can be formed in realistic or cosmologically-motivated theories without additional fermions nor a singular potential.
One of the physically-motivated theories is the AD mechanism \cite{AD}, which includes two types of potentials,
gravity-mediation type and gauge-mediation type. 
The former is described by 
\bea\label{gravity}
&&V_{\rm grav.}(\phi):=\frac{m_{\rm grav.}^2}{2}\phi^2\left[
1+K\ln \left(\frac{\phi}{M}\right)^2
\right]~~  \nonumber  \\
&&{\rm with} ~~ m_{\rm grav.}^2,~M>0,
\eea
while the latter by
\beq\label{gauge}
V_{\rm gauge}(\phi):=m_{\rm gauge}^4 \ln\left(1+\frac{\phi^2}{m_{\rm gauge}^2}\right)~~~
{\rm with} ~~~ m_{\rm gauge}^2>0\ .
\eeq
If we take Maclaurin expansion of the two potentials in the vicinity of $\phi=0$, the latter can be regarded as $V_{4}$ model,  and is inappropriate for our purpose.
Thus, we concentrate on investigating gauged Q-balls in the former potential. 

This paper is organized as follows.
In Sec. II, we show the basic equations of gauged Q-balls. 
In Sec. III, we discuss general properties of ordinary and gauged Q-balls in words of Newtonian mechanics.
In Sec. IV, we review previous results of $V_{4}$ and $V_{\rm V}$ models. 
In Sec. V, we investigate equilibrium solutions in the $V_{\rm grav.}$ model numerically.
Section VI is devoted to concluding remarks.

\section{basic equations}

Consider an SO(2) symmetric scalar field $\bp=(\phi_1,\phi_2)$ coupled to a gauged field $A_\mu$,
\beq\label{S}
{\cal S}=\int d^4x\left[\frac{1}{4}F_{\mu\nu}F^{\mu\nu}
-\frac{1}{2}\eta^{\mu\nu}D_{\mu}\phi_{a} D_{\nu}\phi_{a}-V(\phi) \right],
\eeq
where 
\bea
&&\phi:=\sqrt{\phi_{a}\phi_{a}},~~~
F_{\mu\nu}:=\pa_\mu A_\nu-\pa_\nu A_\mu,\\
&&D_{\mu}\phi_{a}:=\pa_{\mu}\phi_{a}+A_{\mu}\epsilon_{ab}\phi_{b}~(a,b=1,2). 
\eea
To find spherically symmetric and equilibrium solutions with vanishing magnetic fields, we assume
\beq\label{qball-phase}
\bp=\phi(r)(\cos\omega t,\sin\omega t),~~~
A_{0}=A_{0}(r),~~~ A_i=0,
\eeq
where the subscript $i$ denotes spatial components and runs 1 to 3.
Introducing a variable,
\beq
\Omega(r):=\omega+qA_{0}(r),
\eeq
we obtain field equations, 
\bea\label{FEqball}
&&\frac{d^2\phi}{dr^2}+\frac{2}{r}\frac{d\phi}{dr}+\Omega^2 \phi=\frac{dV}{d\phi}, \\
&&\frac{d^2\Omega}{dr^2}+\frac{2}{r}\frac{d\Omega}{dr}=\Omega (q\phi)^{2}. \label{FEqball2}
\eea

The boundary condition we assume is 
\bea
&&{d\phi\over dr}(r=0)=0,~~{d\Omega\over dr}(r=0)=0, \label{BCqball} \\
&&\phi(r\ra\infty)=0,~~\Omega(r\ra\infty)=\omega+\frac{C}{r},  \label{BCqball2}
\eea
where $C$ is a constant.
In numerical calculation we must choose $\Omega$ and $\phi$ at $\tilde{r}=0$ to satisfy the asymptotic conditions 
(\ref{BCqball2}). 
In concrete, we seek for appropriate $\phi (0)$ for a fixed $\Omega (0)$. 

We define the energy and the charge, respectively, as
\bea\label{Edef}
E&=&\int d^3xT_{00}\nn
&=&2\pi \int_0^{\infty}r^2 dr
\left\{\Omega^2\phi^2+\left({d\phi\over dr}\right)^2+\left({d\Omega\over dr}\right)^2+2V\right\},\nn
Q&=&\int d^3x(\phi_1D_0\phi_2-\phi_2D_0\phi_1)\nn
&=&4\pi \int_0^{\infty}r^2\Omega\phi^2dr,
\label{Qdef}\eea
where $T_{00}$ is the time-time component of the energy momentum tensor, which is defined by
\bea
T_{\mu\nu}&=&D_\mu\phi_aD_\nu\phi_a-\eta_{\mu\nu}\left[\frac12(D_\lambda\phi_a)^2+V\right]\nn
&&+F_{\mu\lambda}F_\nu^\lambda-\frac14\eta_{\mu\nu}(F_{\lambda\sigma})^2.
\eea
Equations (\ref{FEqball}), (\ref{FEqball2}) and (\ref{Qdef}) indicate that the sign transformation
$\Omega \to -\Omega$ changes nothing but $Q\to -Q$ with keeping $E$ and $\phi(r)$ unchanged.
Thus, we choose $\Omega >0$ in this paper.

\section{General Properties of Ordinary and Gauged Q-ball Solutions. }

\begin{figure}[htbp]
\psfig{file=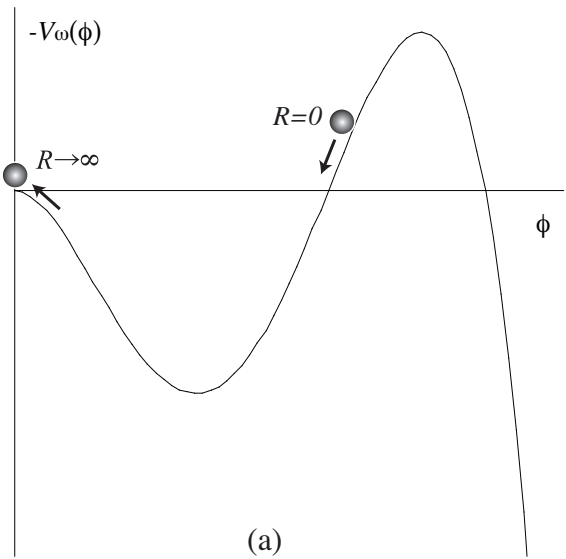,width=3.2in}
\caption{\label{f1}
Interpretation of ordinary Q-balls by analogy with a particle motion in Newtonian mechanics.}
\end{figure}

To begin with, to understand the effect of gauge fields on Q-balls, we review properties of ordinary Q-ball solutions.
The field equations are obtained by putting $\Omega=\omega$=constant in Eq.(\ref{FEqball}),
\beq\label{FEOQ}
\frac{d^2\phi}{dr^2}+\frac{2}{r}\frac{d\phi}{dr}=\frac{dV_\omega}{d\phi},~~~
V_\omega :=V-\frac12\omega^2\phi^2.
\eeq
If one regards the radius $r$ as \lq time\rq\ and the scalar amplitude $\phi(r)$ as \lq the position of a particle\rq,
one can understand solutions in words of Newtonian mechanics, as shown in Fig.\ \ref{f1}.
Equation (\ref{FEOQ}) describes a one-dimensional motion of a particle under the nonconserved force 
due to the effective potential $-V_{\omega}(\phi)$ and the \lq time\rq-dependent friction $-(2/r)d\phi/dr$.
If one chooses the \lq initial position\rq\ $\phi(0)$ appropriately, the static particle begins to roll down the potential slope, climbs up and approaches the origin over infinite time.

From the above picture, one can derive the existing conditions of equilibrium solutions of ordinary Q-balls as follows.
The first condition is that the \lq initial altitude of the particle\rq\  $-V_\omega(\phi(0))$ is larger than the \lq final altitude\rq\
$-V_\omega(\phi(\infty))=0$, which leads to
\beq\label{cond1}
{\rm max}[-V_\omega(\phi)]>0,~~i.e.,~~
{\rm min}\left[{2V\over\phi^2}\right]<\omega^2.
\eeq
The second condition is that the \lq particle climbs up\rq\ at $r\to\infty$, which leads to
\beq\label{cond2}
\lim_{\phi\to +0}\frac{1}{\phi}\left(-{dV_\omega\over d\phi}\right)=
\lim_{\phi\to +0}\frac{1}{\phi}\left(\omega^2\phi-{dV\over d\phi}\right)<0\ .
\eeq
If the lowest-order term of $V$ is quadratic, i.e., $\ds V=\frac12m^2\phi^2+O(\phi^3)$, the second condition (\ref{cond2}) reduces to
\beq\label{cond22}
\omega^2<m^2={d^2V\over d\phi^2},
\eeq
which gives the upper limit of $\omega^2$.
The conditions (\ref{cond1}) and (\ref{cond22}) were originally obtained by Coleman \cite{Col85}).

We should not, however, apply the second condition (\ref{cond22}) to the cone-shape potential $V_{\rm V}$ in (\ref{VV}) nor the AD gravity-mediation type $V_{\rm grav.}$ in (\ref{gravity}) because their lowest-order term is not quadratic. Instead we should go back to the condition (\ref{cond2}). In the case of $V_{\rm V}$, if we take $\lambda>0$, the condition (\ref{cond2}) is satisfied regardless of $\omega$. Similarly, in the case of $V_{\rm grav.}$, if we take $K<0$, the condition (\ref{cond2}) is satisfied regardless of $\omega$. 

Now let us move on to gauged Q-balls. Without specifying a potential $V$, we can show that $\Omega^2$ is a 
monotonically increasing function of $r$~\cite{Arodz}.
Using a variable $\ds f:=r^2\frac{d\Omega}{dr}$, we can rewrite Eq. (\ref{FEqball2}) as
\bea\label{FEqball2-2}
&&\frac{df}{dr}=\Omega (qr\phi)^{2},~~~
\frac{d\Omega}{dr}={f\over r^2}\ .
\eea
The Taylor expansion of $\Omega$ and $f$ up to the first order is expressed as
\bea
f(r+\Delta r)&=&f(r_0)+(qr_0\phi(r))^2\Omega(r)\Delta r+O(\Delta r^2),\nn
\Omega(r+\Delta r)&=&\Omega(r)+{f(r)\over r^2}\Delta r+O(\Delta r^2).
\label{Taylor}\eea
By definition $f(0)=0$. If $\Omega (0)>0$, then $f(\Delta r)>0$. Equation (\ref{Taylor}) indicates that at every step $r\to r+\Delta r$ both $f$ and $\Omega$ increases. 
Similarly, if $\Omega (0)<0$, then $f$ and $\Omega$ decreases at every step.
Thus we can conclude that  $\Omega^2$ is a monotonically increasing function of $r$.

\begin{figure}[htbp]
\psfig{file=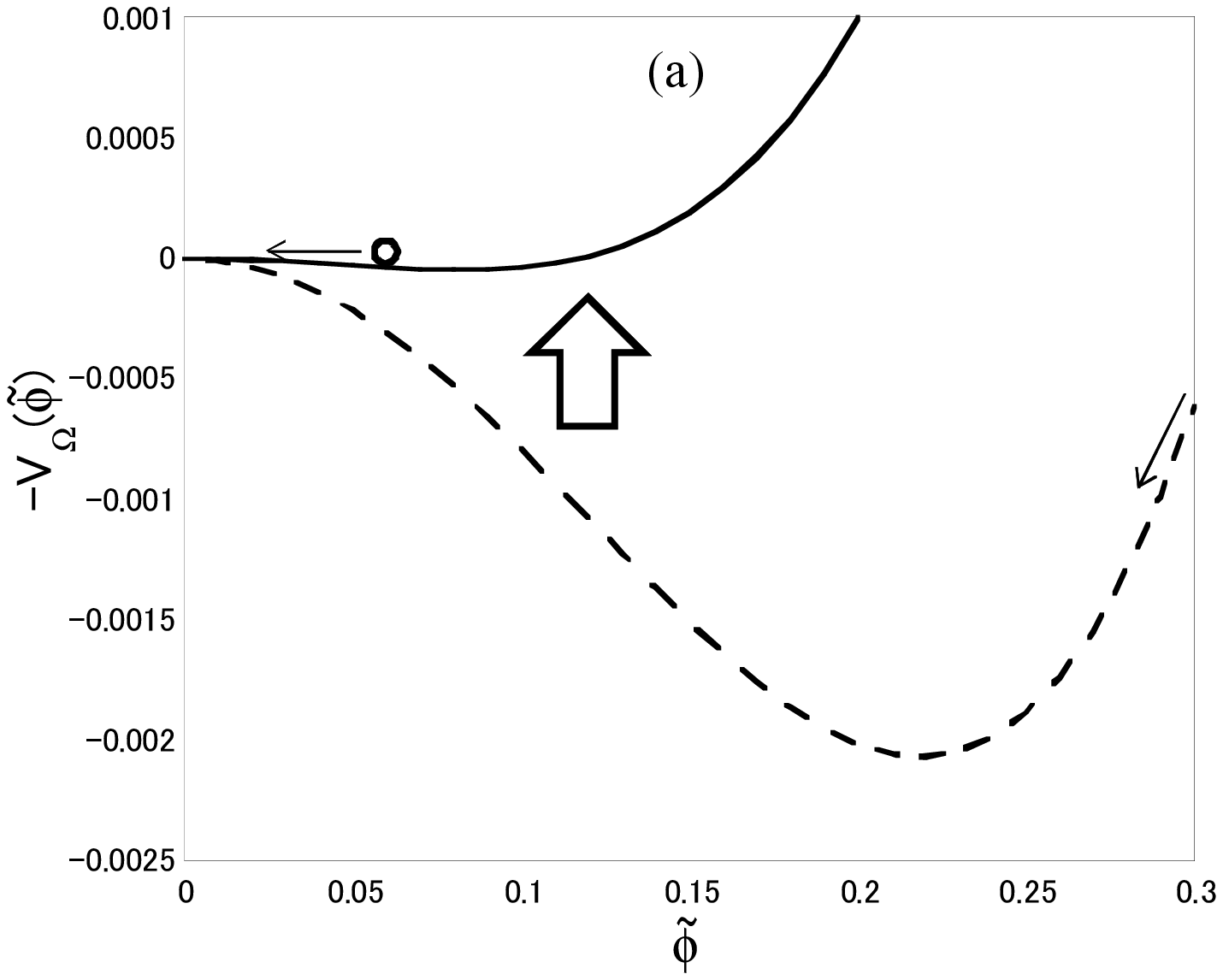,width=3.2in}
\psfig{file=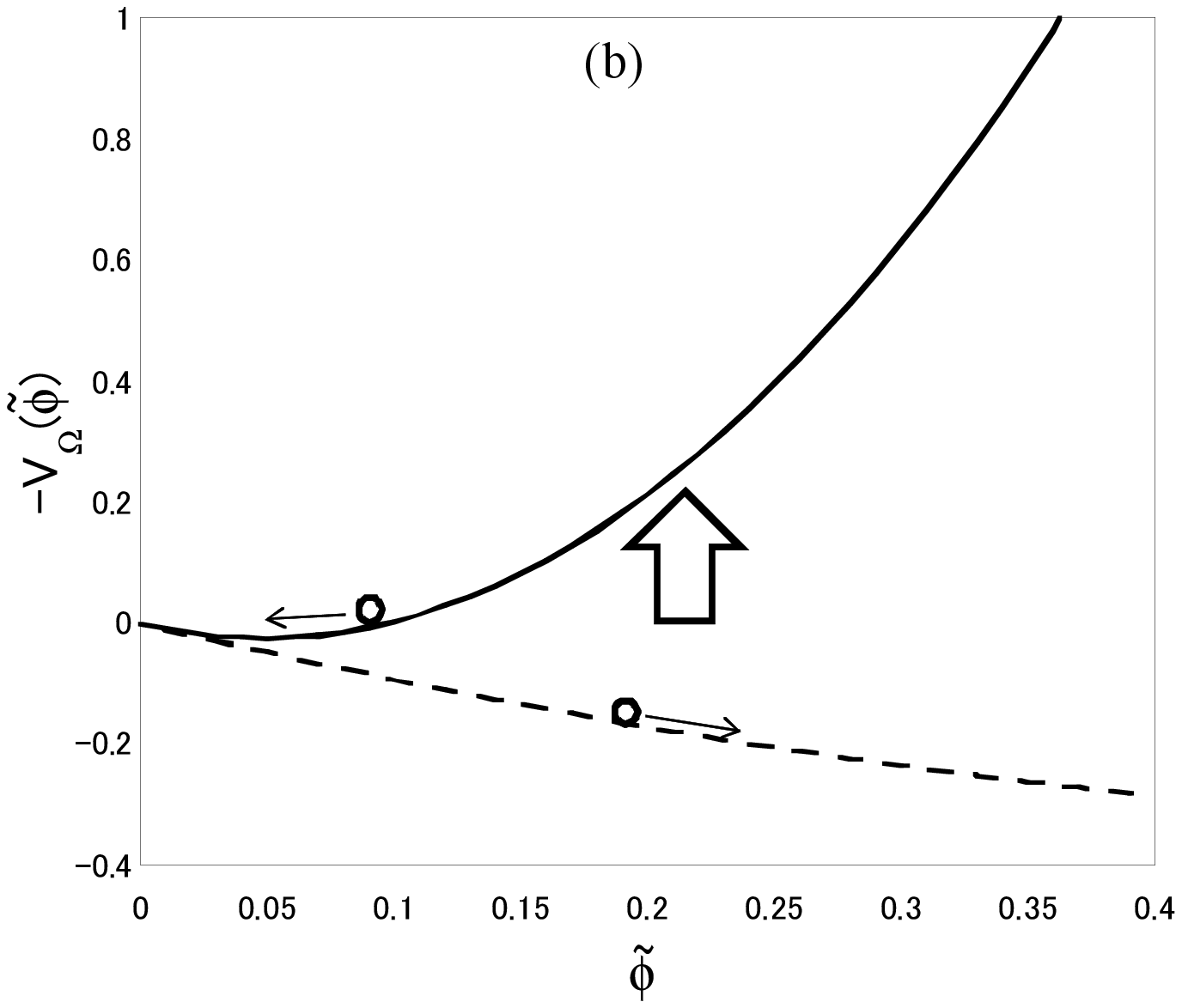,width=3.2in}
\caption{\label{Newton}
Interpretation of gauged Q-balls by analogy with a particle motion in Newtonian mechanics. Examples of
(a) monotonic solutions in $V_{4}$ model and (b) nonmonotonic solutions in $V_{\rm V}$ model.}
\end{figure}

We can interpret their equilibrium solutions in words of Newtonian mechanics in the same fashion, except that the potential of a particle is \lq time\rq-dependent,
\beq
V_{\Omega}=V-\frac12\Omega^2\phi^2.
\eeq
Because the \lq potential energy of the particle\rq\ $-V_\Omega$ increases as the \lq time\rq\ $r$ increases,
the \lq initial altitude\rq\ $-V_\Omega(0)$ is not necessarily larger than the \lq final altitude\rq\ $-V_\Omega(\infty)=0$, that is,
there is no condition which corresponds to (\ref{cond1}).
However, the condition that the \lq particle climbs up\rq\ at $r\to\infty$ should hold, we find an existing condition, which corresponds to (\ref{cond2}),
\beq\label{gaugedQexist}
\lim_{\phi\to +0}\frac{1}{\phi}\left(-{dV_\Omega\over d\phi}\right)=
\lim_{\phi\to +0}\frac{1}{\phi}\left(\Omega^2\phi-{dV\over d\phi}\right)<0\ .
\eeq

Figure \ref{Newton} illustrates the  \lq time-dependent potential of a fictitious particle \rq $-V_\Omega$.
As $r$ increases, $\Omega^2$ also increases; then $-V_\Omega$ goes up as shown in the figure.
There are two types of solutions.
One is monotonic solutions as shown in (a): $\phi$ decreases monotonically as $r$ increases.
The other is nonmonotonic solutions as shown in (b): $\phi$ increase initially, but after the sign of $dV_\Omega/d\phi$ changes, $\phi$ turns to decreases.
The latter type exposes a characteristic of gauged Q-balls, which appears in the $V_{\rm V}$ and $V_{\rm grav.}$ models.

\begin{figure}[htbp]
\psfig{file=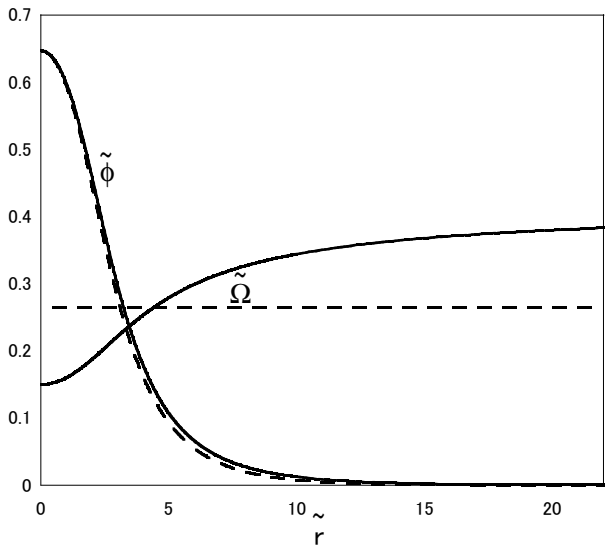,width=3.2in}
\caption{\label{V4m02Q09r-phi}
The field configurations of $\tilde{\phi}$ and $\tilde{\Omega}$ for the $V_{4}$ model with $\tilde{m}^{2}=0.2$ 
and $\tilde{Q}=9$. The dashed and solid lines correspond to the ordinary and gauged Q-balls, respectively.}
\end{figure}
\section{Review of previous results}

In this section we review gauged Q-ball solutions in the $V_{4}$ model \cite{Lee} and in the $V_{\rm V}$ model \cite{Arodz}.

\subsection{$V_{4}$ model}

For the $V_4$ model (\ref{V4}), the necessary condition of existing equilibrium solutions (\ref{gaugedQexist}) is expressed as 
\beq\label{gaugedQexist2}
\lim_{r\to\infty}\Omega^2 <m^2.
\eeq
Because $\Omega^2$ is an increasing function of $r$,  the condition (\ref{gaugedQexist2}) would give a rather strong constraint on the parameter range of existing equilibrium solutions.

We confirm this expectation by numerical calculation as follows.
We rescale the quantities as 
\bea
&&\tp:=\frac{q\phi}{\sqrt{\lambda}M},~~\tilde{\Omega}:=\frac{\Omega}{\sqrt{\lambda}M},~~ \nonumber  \\
&&\tilde{r}:= \sqrt{\lambda}Mr, ~~\tilde{m}:= \frac{m}{\sqrt{\lambda}M},  \nonumber  \\
&&\tilde{Q}:= q^2 Q, ~~\tilde{E}:= \frac{q^2 E}{\sqrt{\lambda}M}.
\label{rescale-V4}
\eea
In Fig.~\ref{V4m02Q09r-phi}, as an example, we show the solution with $\lambda=q=1$, $\tilde{m}^{2}=0.2$ and $\tilde{Q}=9$. 
The dashed and solid lines correspond to the ordinary and gauged Q-balls, respectively.
In this solution, the distributions of the scalar field $\tilde{\phi}(\tilde r)$ almost coincide; however,
because $\tilde{\Omega}$ increases as a function of $\tilde{r}$, 
the condition (\ref{gaugedQexist2}) is narrowly satisfied. 
Actually, Fig.~\ref{Newton} (a) shows the effective potential of this case.
Equation (\ref{FEqball2}) tells us that, in order for $\tilde{\Omega}$ to be small in the asymptotic region, 
$\tilde{r}\tilde{\phi}$ must also be small there; 
this indicates that $\tilde{Q}$ has an upper limit. 

\subsection{$V_{\rm V}$ model}

Because the $V_{\rm V}$ model (\ref{VV}) has a linear term, the condition (\ref{gaugedQexist}) is satisfied if $\lambda>0$.
Contrary to the case of the $V_4$ model, this condition does not put any restriction on $\Omega$. 
Therefore, large gauged Q-balls are expected in this model.

\begin{figure}[htbp]
\psfig{file=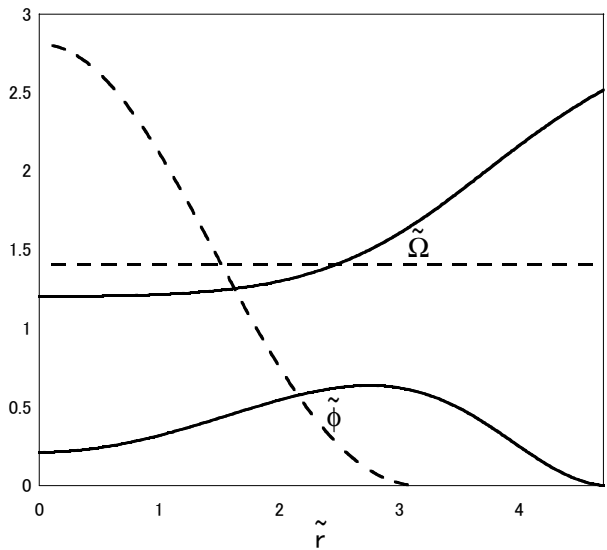,width=3.2in}
\caption{\label{VVQ120r-phi}
The field configurations of $\tilde{\phi}$ and $\tilde{\Omega}$ for the $V_{\rm V}$ model with $\tilde{Q}=120$. 
The dashed and solid lines correspond to the ordinary and gauged Q-balls, respectively.}
\end{figure}

Using the normalized coupling $\kappa:={q\lambda}/{\sqrt{2}}$, we rescale the quantities as 
\bea
&&\tp:=\frac{q\phi}{\sqrt{\kappa}},~~\tilde{\Omega}:=\frac{\Omega}{\sqrt{\kappa}},~~
\tilde{r}:= \sqrt{\kappa}r, \nonumber  \\
&&\tilde{Q}:= q^2 Q, ~~\tilde{E}:= \frac{q^2 E}{\sqrt{\kappa}}.
\label{rescale-VV}
\eea
In Fig.~\ref{VVQ120r-phi}, we show the field configurations of $\tilde{\phi}$ and $\tilde{\Omega}$ 
with $\tilde{Q}=120$. 
The dashed and solid lines correspond to the ordinary and gauged Q-balls, respectively. 
In the case of gauged Q-balls, $\tilde{\phi}$ initially increases as a function of $\tilde{r}$ and 
takes a maximum value at $\tilde{r}=\tilde{r}_{\rm max}\neq 0$; then it decreases due to the increase of $\tilde{\Omega}$.
This behavior can be understood by the effective potential shown in Fig.~\ref{Newton} (b). 
Here we have defined $\tilde{r}_{\rm max}$ as the value of $\tilde{r}$ where $\tilde{\phi}$ takes a maximum value.
In the case of ordinary Q-balls, by contrast, $\tilde{r}_{\rm max}$ is always zero.

\begin{figure}[htbp]
\psfig{file=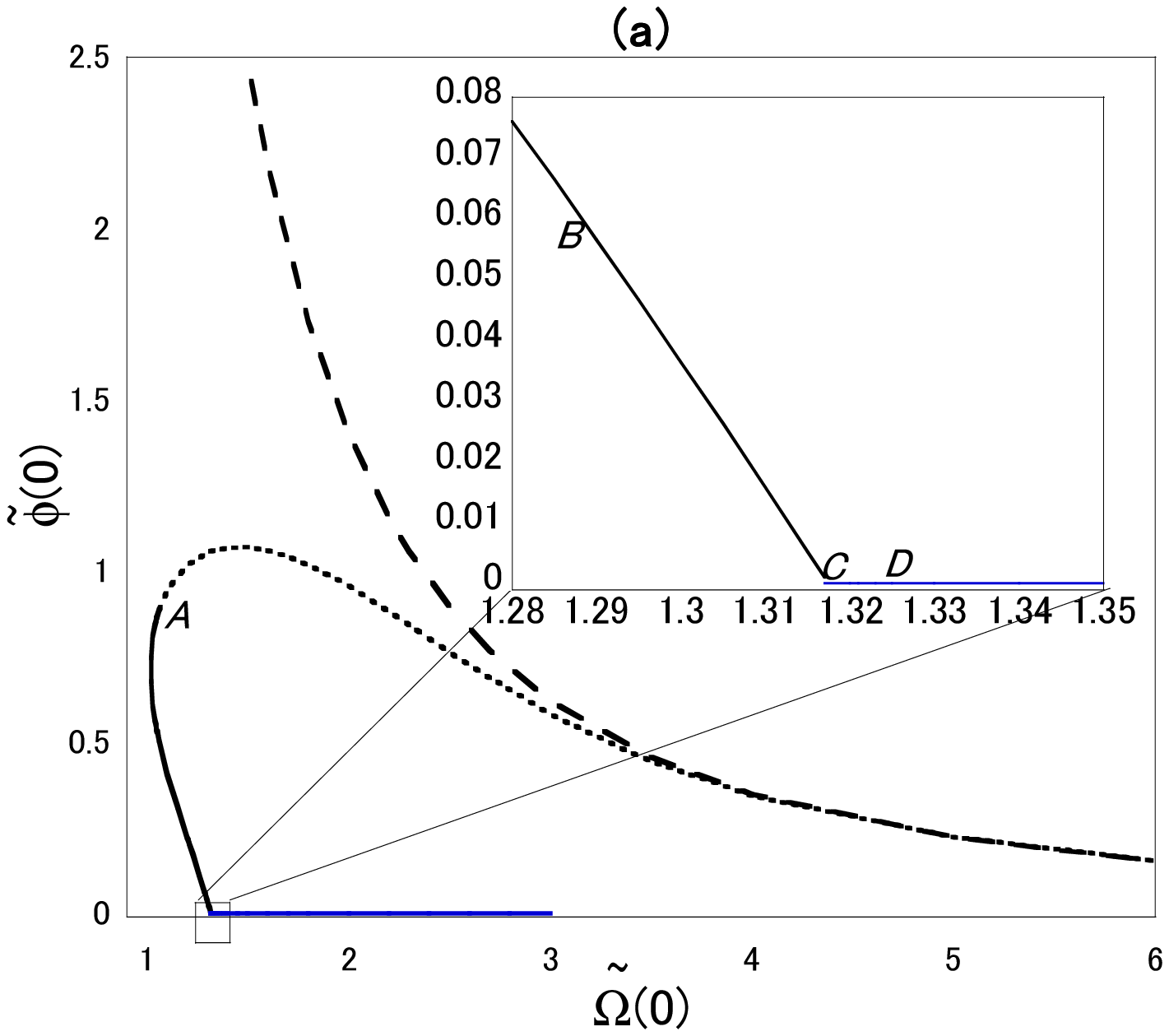,width=3.2in}
\psfig{file=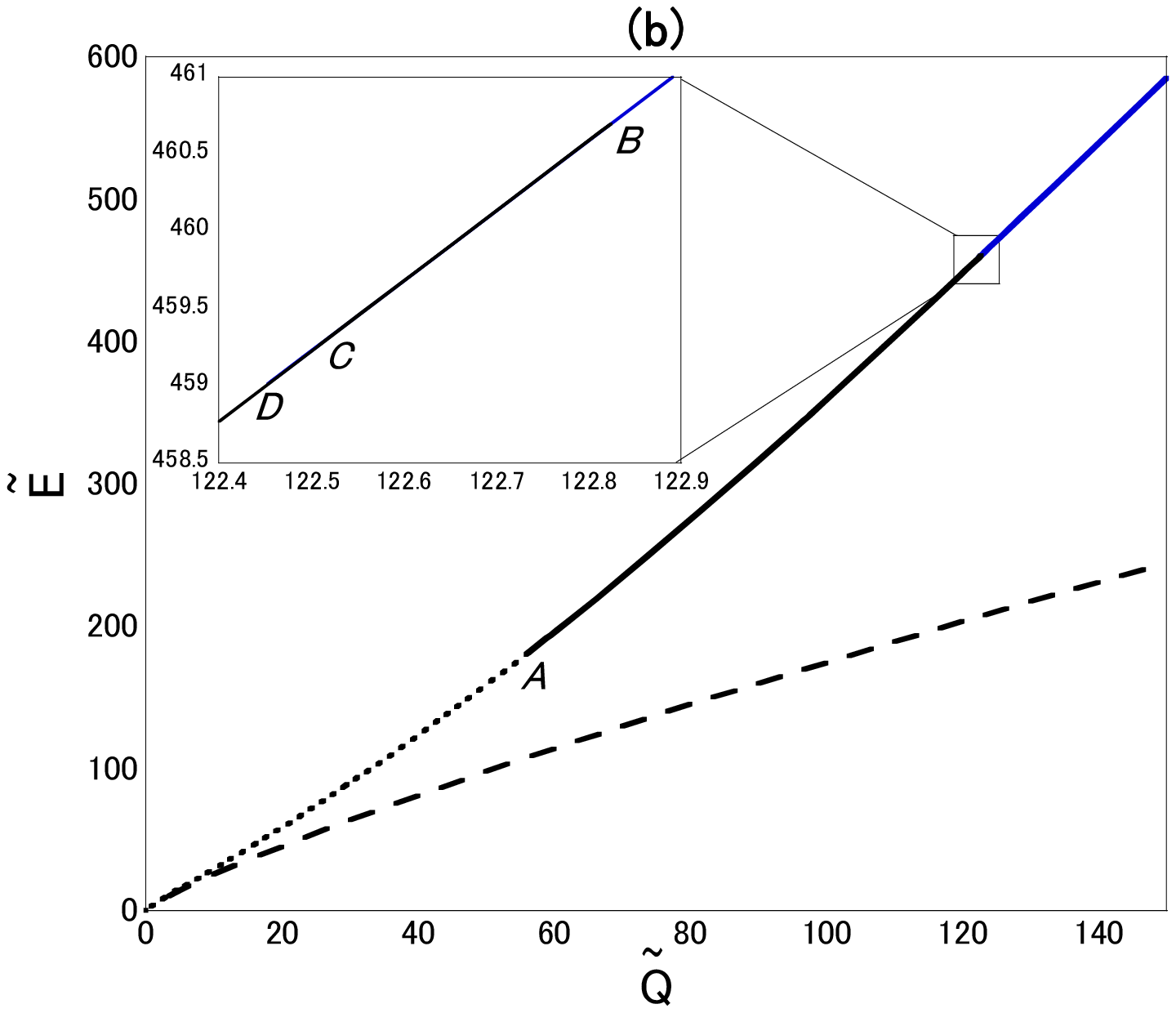,width=3.2in}
\caption{\label{Omega-phiVV}
(a) $\tilde{\Omega}(0)$-$\tilde{\phi}(0)$ and (b) $\tilde{Q}$-$\tilde{E}$ relations for the $V_{\rm V}$ model. 
The dashed line corresponds to the ordinary Q-balls. 
The dotted and black solid lines correspond to the gauged Q-balls with $\tilde{r}_{\rm max}= 0$ and those with 
$\tilde{r}_{\rm max}\neq 0$, respectively.
Blue solid line corresponds to the Q-shell solutions. }
\end{figure}

We show the $\tilde{\Omega}(0)$-$\tilde{\phi}(0)$ and $\tilde{Q}$-$\tilde{E}$ relations in Fig.~\ref{Omega-phiVV} (a) and (b), respectively.
The dashed line corresponds to the ordinary Q-balls. 
The dotted and black solid lines correspond to the gauged case with 
$\tilde{r}_{\rm max}= 0$ and that with $\tilde{r}_{\rm max}\neq 0$, respectively. 
Blue solid line corresponds to the Q-shell solutions that will be explained below. 

In the case of ordinary Q-balls ($\tilde{\Omega}=\tilde{\omega}$), the $\tilde{\Omega}(0)$-$\tilde{\phi}(0)$ relation,  
which was represented by the dashed line in (a), can be understood as follows.
In the picture of a particle motion in Newtonian mechanics, which was shown in Fig.\ \ref{f1}, if we ignore the ``nonconserved force" term, $(2/r)d\phi/dr$, the maximum of $\tilde{\phi}$, $\tilde{\phi}_{\rm max}=\tilde{\phi}(0)$ is determined by the nontrivial solution of $V_{\Omega}=0$. Then we obtain
\bea
\tilde{\phi}(0)=\frac{2}{\tilde{\Omega}^2}, 
\label{phimax-VV}
\eea
which approximates the dashed line in (a).

In the case of gauged Q-balls, the $\tilde{\Omega}(0)$-$\tilde{\phi}(0)$ relation for large 
$\tilde{\Omega}(0)$ (small $\tilde{Q}$), which is represented by the dotted line in (a), almost coincides with that for ordinary Q-balls.
For small $\tilde{\Omega}(0)$ (large $\tilde{Q}$), however, the $\tilde{\Omega}(0)$-$\tilde{\phi}(0)$ relation for ordinary Q-balls and that for gauged Q-balls are qualitatively different.
Nevertheless, it is surprising that there is no qualitative difference  in $\tilde{Q}$-$\tilde{E}$ relation between
solutions with $\tilde{r}_{\rm max}= 0$ and those with $\tilde{r}_{\rm max}\neq 0$.
Both solutions are on the same quasi-linear relation across the point $A$. 

$Q$ reaches a maximum at the point $B$ where cusp structure appears in the $\tilde{Q}$-$\tilde{E}$ plane.
Q-ball solutions with the boundary conditions (\ref{BCqball}) disappear at the point $C$ where $\tilde{\phi}(0)\to 0$. 
However, Arod\'z and Lis \cite{Arodz} found a new type of solutions with boundary conditions (\ref{BCqball2}) and
\bea
&&\phi (r)={d\phi\over dr}(r)={d\Omega\over dr}(r)=0,~~{\rm for}~0<r<r_{0},  \label{BCqshell}
\eea
which are different from (\ref{BCqball}), and called them ``Q-shells."
The $\tilde{Q}$-$\tilde{E}$ curve of Q-shells is smoothly connected to that of Q-balls at the point $C$.
As $\tilde{\Omega}(0)$ increases, $Q$ decreases and reaches a minimum at another cusp $D$ in the $\tilde{Q}$-$\tilde{E}$ plane; then $Q$ turns to increase without upper limit.
If we magnify Fig.\ \ref{Omega-phiVV}(b) further, we see that the solutions $B$-$C$-$D$ have slightly larger values of $\tilde{E}$ than those of the other solutions with the same $Q$. 
If we apply catastrophe theory~\cite{PS78}, we find that the solution sequence $B$-$C$-$D$ is unstable, while the other two sequences are stable and crosses in the $\tilde{Q}$-$\tilde{E}$ plane.

\section{the AD mechanism for gravity mediation}

As we discussed in the previous section, to obtain large $\tilde{Q}$ solutions, $\tilde{\Omega}$ should become so large without violating the condition (\ref{gaugedQexist}).
$\tilde{\Omega}$ is not constrained by (\ref{gaugedQexist}) at all if 
\beq
\lim_{\phi\ra +0}{dV\over d\phi}>0.
\eeq
Because the AD gravity mediation model (\ref{gravity}) with $K<0$ satisfies this condition, we can expect that 
it allows for large $Q$ solutions. This special property is in common with the V-shaped model.

We rescale the quantities in (\ref{gravity}) as 
\bea
&&\tp:=\frac{q\phi}{M},~~\tilde{\Omega}:=\frac{\Omega}{M},~~ \nonumber  \\
&&\tilde{r}:= Mr, ~~\tilde{m}_{\rm grav.}:= \frac{m_{\rm grav.}}{M},\nonumber  \\
&&\tilde{Q}:= q^2 Q, ~~\tilde{E}:= \frac{q^2 E}{M}.
\label{rescale-gravity}
\eea
We fix $\tilde{m}_{\rm grav.}=q=1$ below. 

\begin{figure}[htbp]
\psfig{file=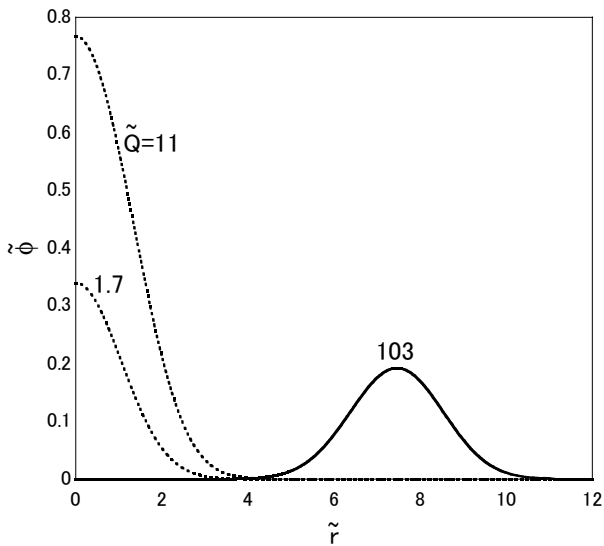,width=3.2in}
\caption{\label{K-1fields}
The field configurations of $\tilde{\phi}$ for gauged Q-balls with $K =-1$ and  $\tilde{Q}\simeq 1.7$, $11$ and $103$.
}
\end{figure}

We show some solutions of gauged Q-balls in Fig.~\ref{K-1fields}; we choose $K =-1$ and obtain 
solutions with $\tilde{Q}=1.7$ and $11$, in which case $r_{\rm max}=0$, and that 
with $\tilde{Q}=103$, in which case $r_{\rm max}\ne0$. 
As $\tilde{Q}$ increases, the field configuration becomes shell-like and the location of the shell becomes farther from the center.
This behavior is explained by repulsive Coulomb force of electric charge.
These configurations are just like ``Q-shells," which were obtained by Arod\'z and Lis for the V-shaped model \cite{Arodz}.
The difference is that we use the boundary condition (\ref{BCqball}) and (\ref{BCqball2}) consistently and 
give tiny but nonzero value for $\tilde{\phi}(0)$, while they adopted the special boundary condition (\ref{BCqshell}).

\begin{figure}[htbp]
\psfig{file=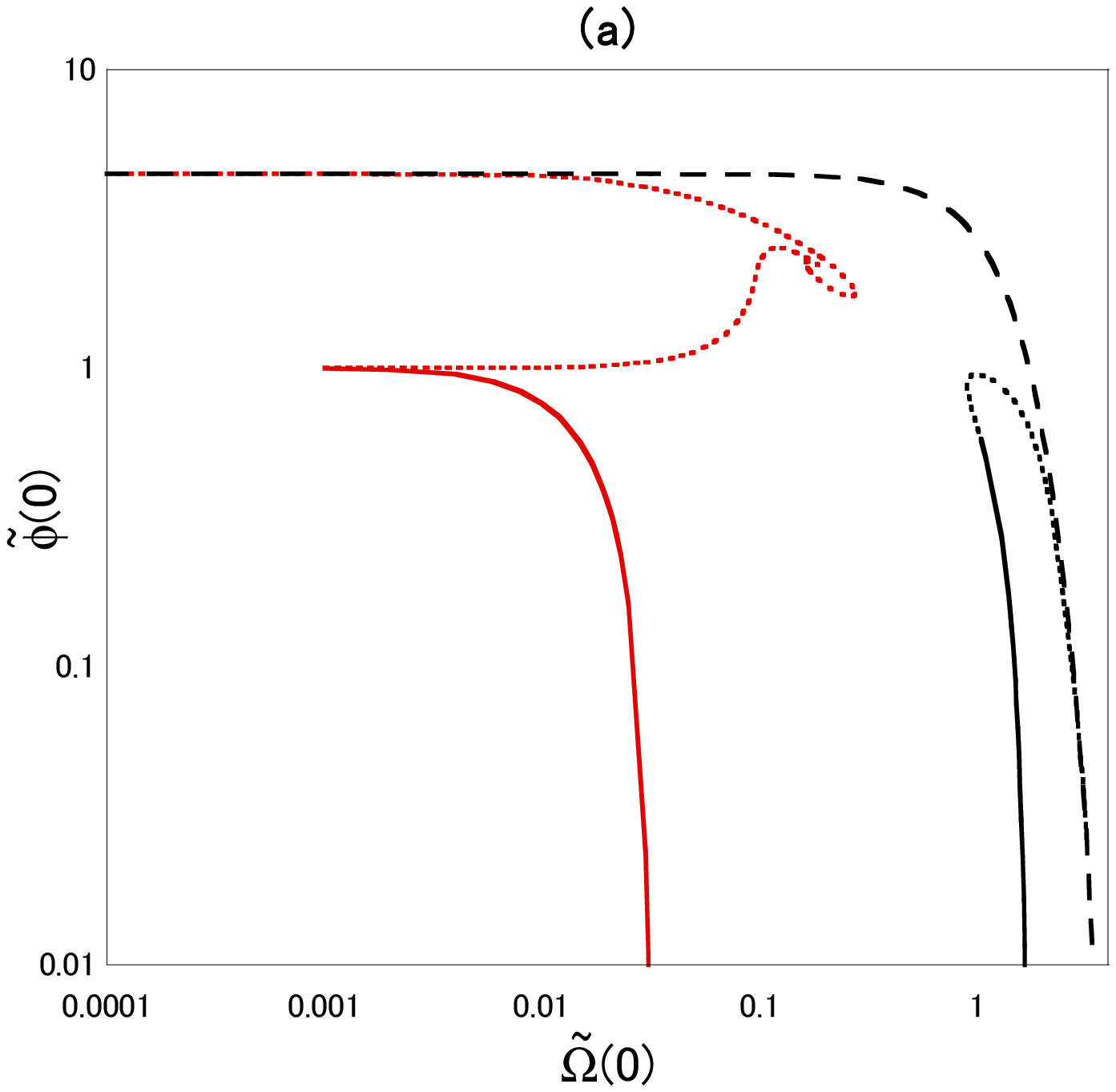,width=3.2in}
\psfig{file=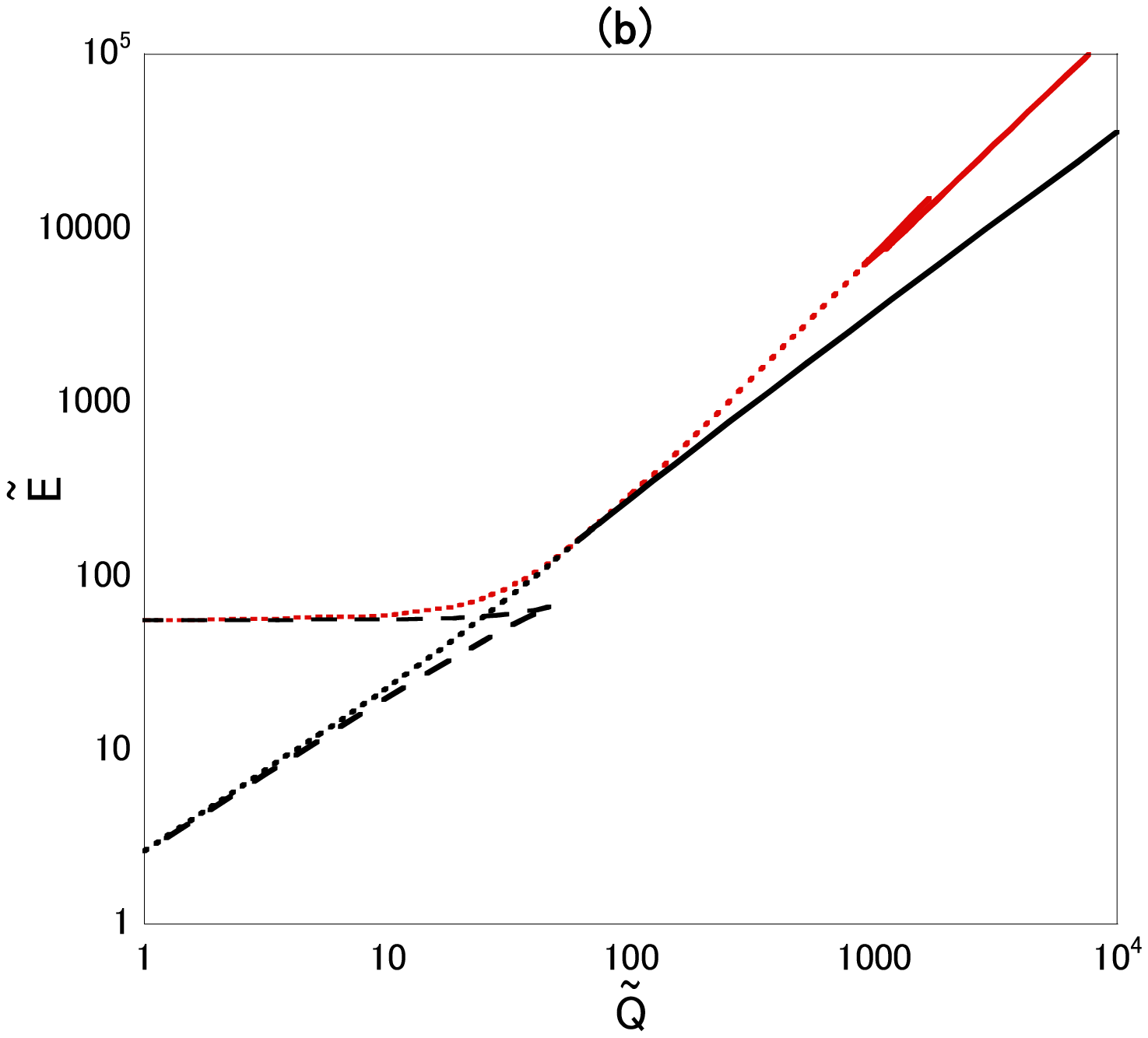,width=3.2in}
\caption{\label{K-1}
(a) $\tilde{\Omega}(0)$-$\tilde{\phi}(0)$ and (b) $\tilde{Q}$-$\tilde{E}$ relations for $K =-1$.
The dashed lines correspond to ordinary Q-balls.
The dotted and solid lines correspond to gauged Q-balls with $\tilde{r}_{\rm max}= 0$ and those with 
$\tilde{r}_{\rm max}\neq 0$, respectively. 
}
\end{figure}

We show the $\tilde{\Omega}(0)$-$\tilde{\phi}(0)$ and $\tilde{Q}$-$\tilde{E}$ relations for $K =-1$ 
in Fig.~\ref{K-1}. For reference, we also plot the relations for ordinary Q-balls ($\Omega=\omega$), which 
are represented by the dashed lines.
Their extreme behavior in the thin-wall limit ($\omega\ra\infty$) and in the thick-wall limit ($\omega\ra0$) can be discussed analytically as follows \cite{TS2}.
The maximum of $\phi$,  $\tilde{\phi}_{\rm max}=\tilde{\phi}(0)$, can be estimated by the nontrivial solution of $V_{\Omega}=0$:
\bea
\tilde{\phi}_{\rm max}=e^{\frac{1-\tilde{\omega}^2}{-2K}}. 
\label{phimax-AD}
\eea
Because the energy and the charge are roughly estimated as
\beq
E\sim V(\phi_{\rm max})R^3,~~~
Q\sim\omega\phi_{\rm max}^{~~~2}R^3,
\eeq
where $R$ is the typical radius, we find
\bea
\omega\ra0&:&\phi_{\rm max}\ra{\rm nzf},~~~E\ra{\rm nzf},~~~Q\ra0,\nn
\omega\ra\infty&:&\phi_{\rm max}\ra0,~~~E\ra0,~~~Q\ra0,
\eea
where nzf denotes nonzero finite. Therefore, there is an upper limit $Q_{\rm max}$. 
This analytic estimate agrees with the numerical results in Fig.\ \ref{K-1}. 
There are two sequences of solutions which merge at the cusp. 
We suppose by energetics that the sequences with high energy are unstable (unstable branch) while 
those with low energy stable (stable branch).

\begin{figure}[htbp]
\psfig{file=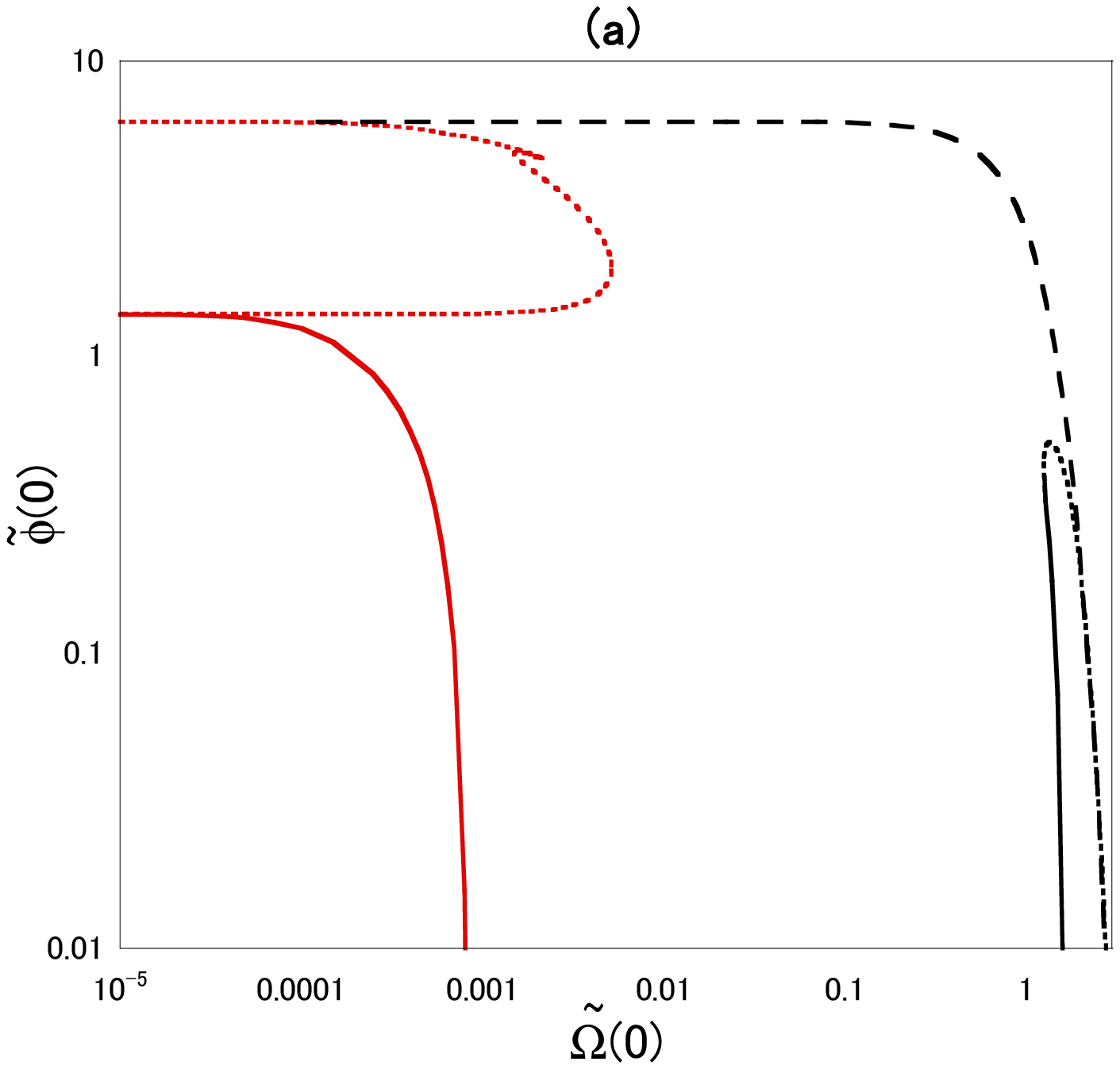,width=3.2in}
\psfig{file=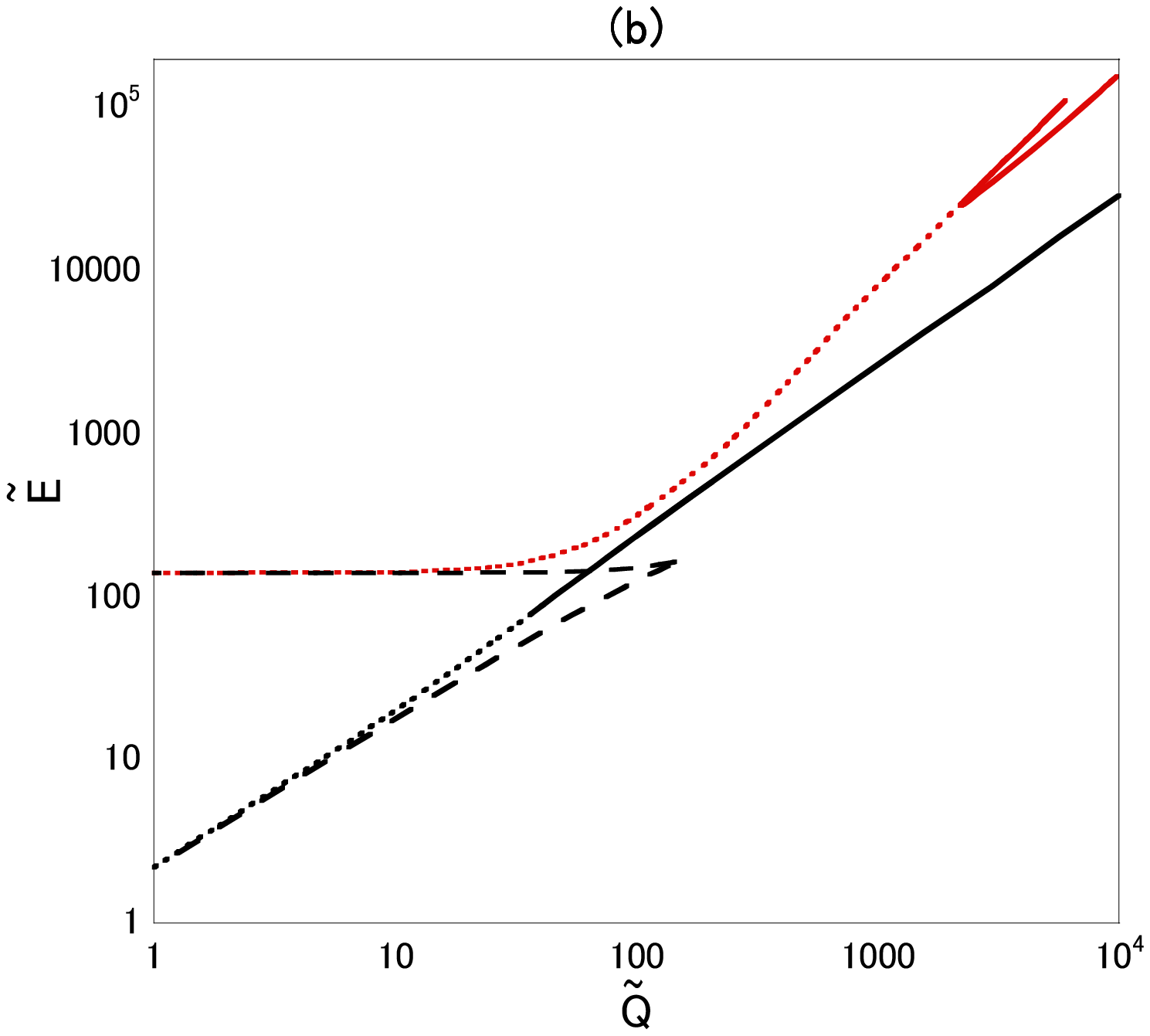,width=3.2in}
\caption{\label{K-06}
(a) $\tilde{\Omega}(0)$-$\tilde{\phi}(0)$ and (b) $\tilde{Q}$-$\tilde{E}$ relations for $K =-0.6$.
}
\end{figure}
\begin{figure}[htbp]
\psfig{file=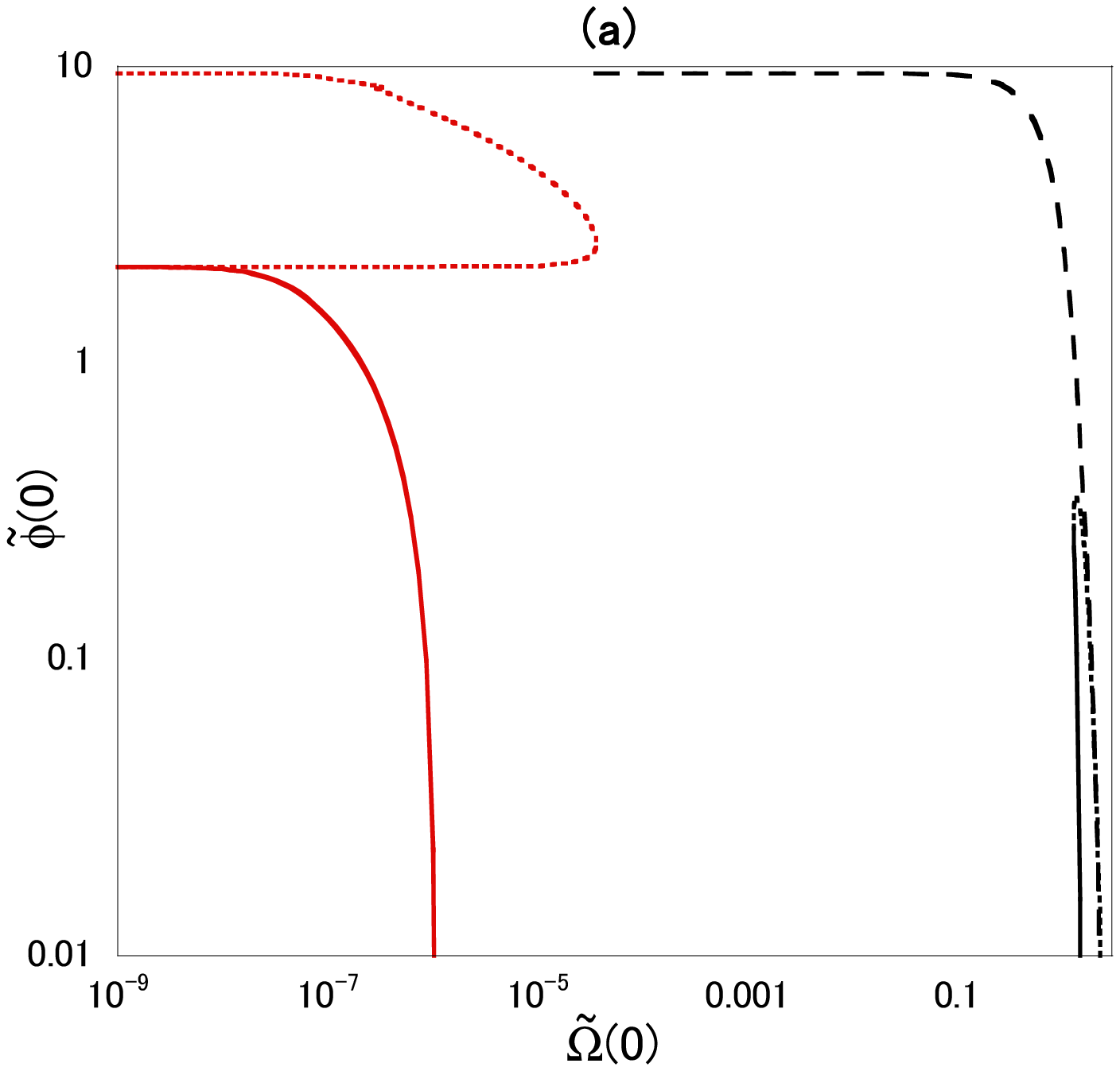,width=3.2in}
\psfig{file=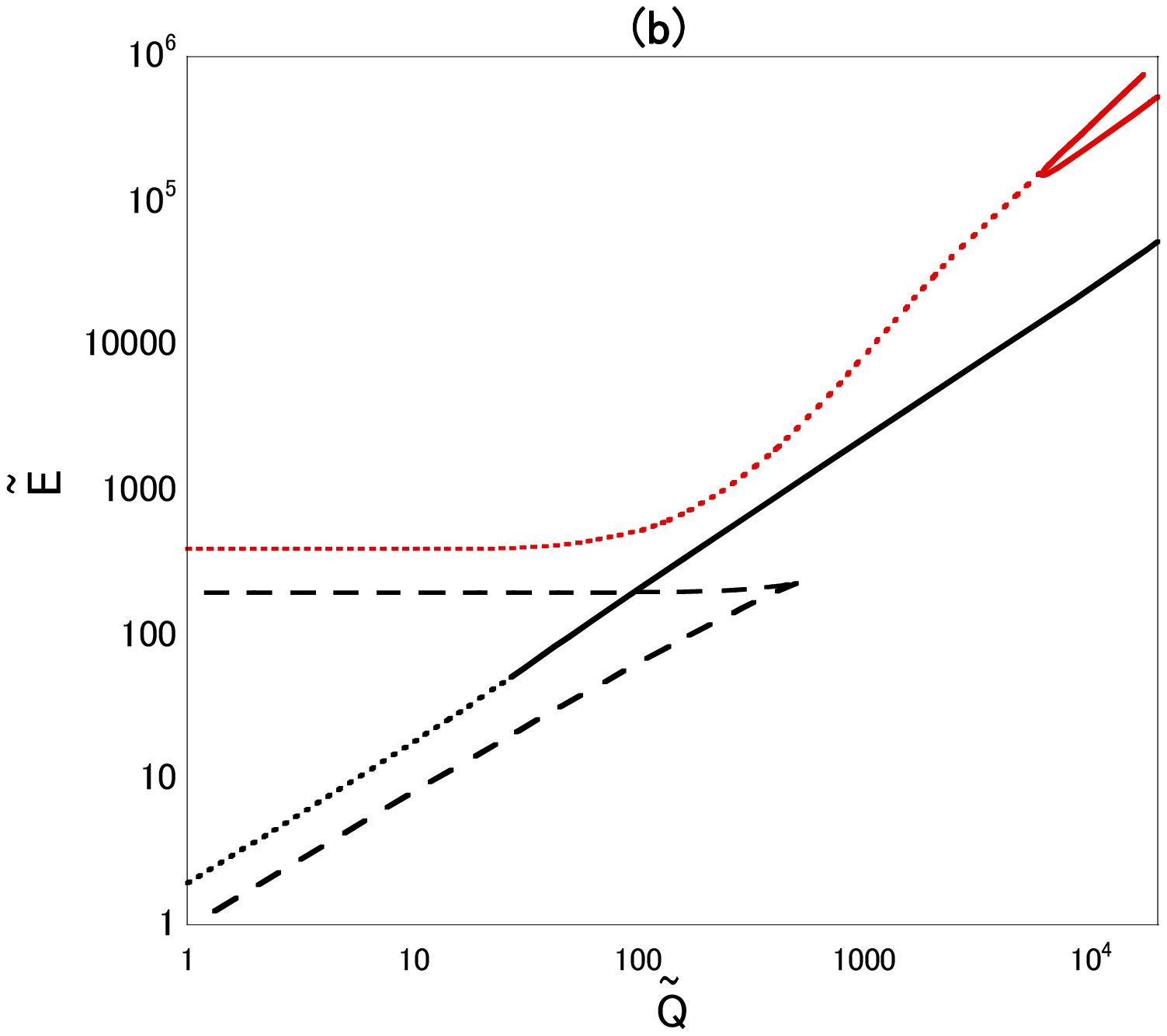,width=3.2in}
\caption{\label{K-04}
(a) $\tilde{\Omega}(0)$-$\tilde{\phi}(0)$ and (b) $\tilde{Q}$-$\tilde{E}$ relations for $K =-0.4$. 
}
\end{figure}

The results for gauge Q-balls are represented by the dashed lines ($\tilde{r}_{\rm max}= 0$) and 
the solid lines ($\tilde{r}_{\rm max}\neq 0$). 
The solutions denoted by red lines correspond to those with small $\omega$ and unstable branch, 
while those by black lines large $\omega$ and stable branch. 
For dotted lines, the gauged Q-balls are  similar to the ordinary Q-balls (dashed lines). In contrast, 
due to the nonmonotonic behavior of $\tilde{\phi}(\tilde{r})$ (i.e., $\tilde{r}_{\rm max}\neq 0$), the 
properties of gauged Q-balls with solid lines and ordinary Q-balls are quite different. 

As for the stable solutions denoted by the black lines, both $\tilde{\Omega}(0)$-$\tilde{\phi}(0)$ 
and $\tilde{Q}$-$\tilde{E}$ relations of solutions are similar to those of the $V_{\rm V}$ model, except that cusp structure does not appear in the $\tilde{Q}$-$\tilde{E}$ plane in 
Fig.~\ref{K-1}(b).
Because $\tilde{E}$ is a monotonically increasing function of $\tilde{Q}$ we judge that all equilibrium solutions by black lines are stable. We also suppose by energetics that the solutions denoted by red lines are unstable. 

\begin{figure}[htbp]
\psfig{file=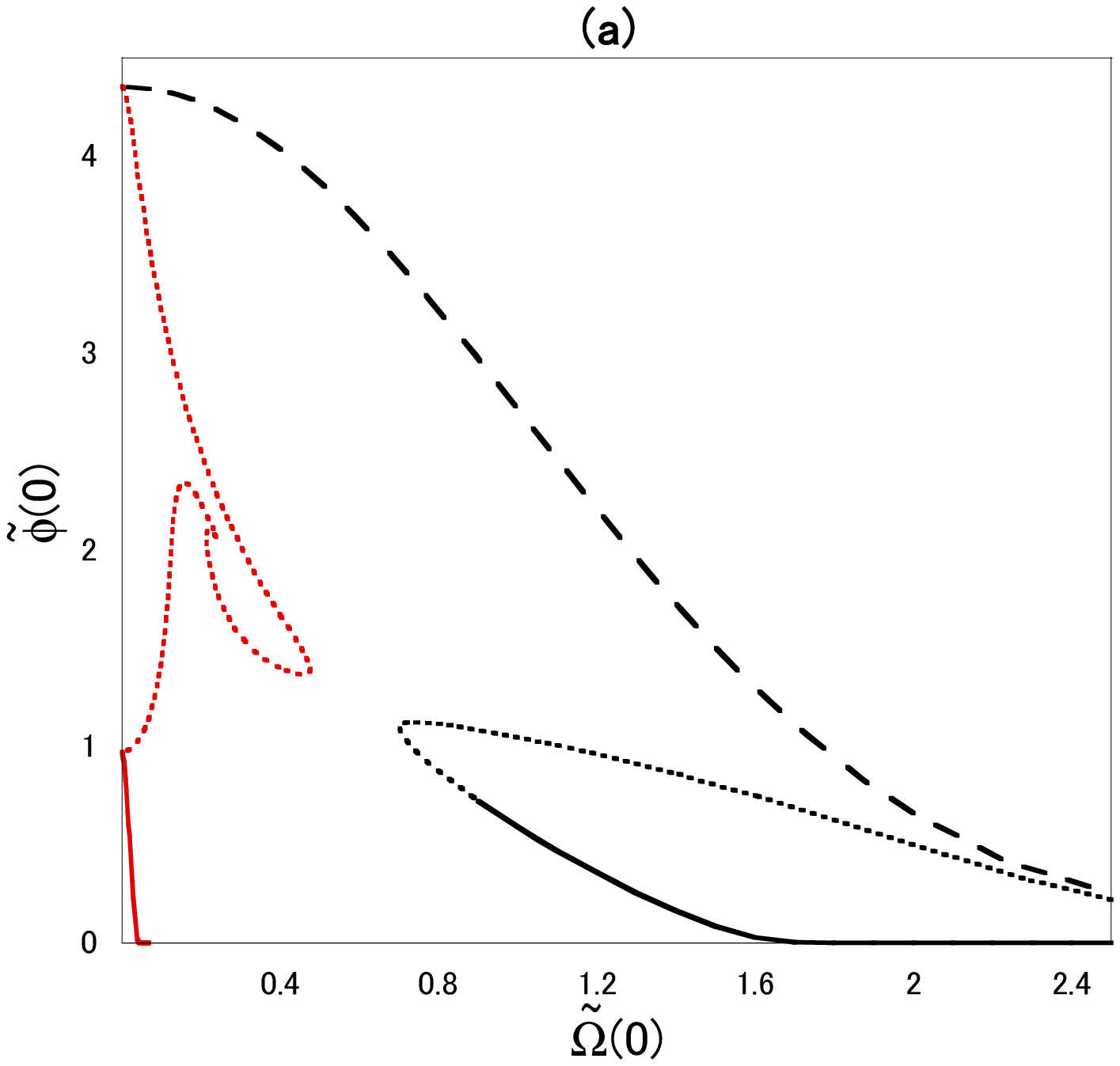,width=3.2in}
\psfig{file=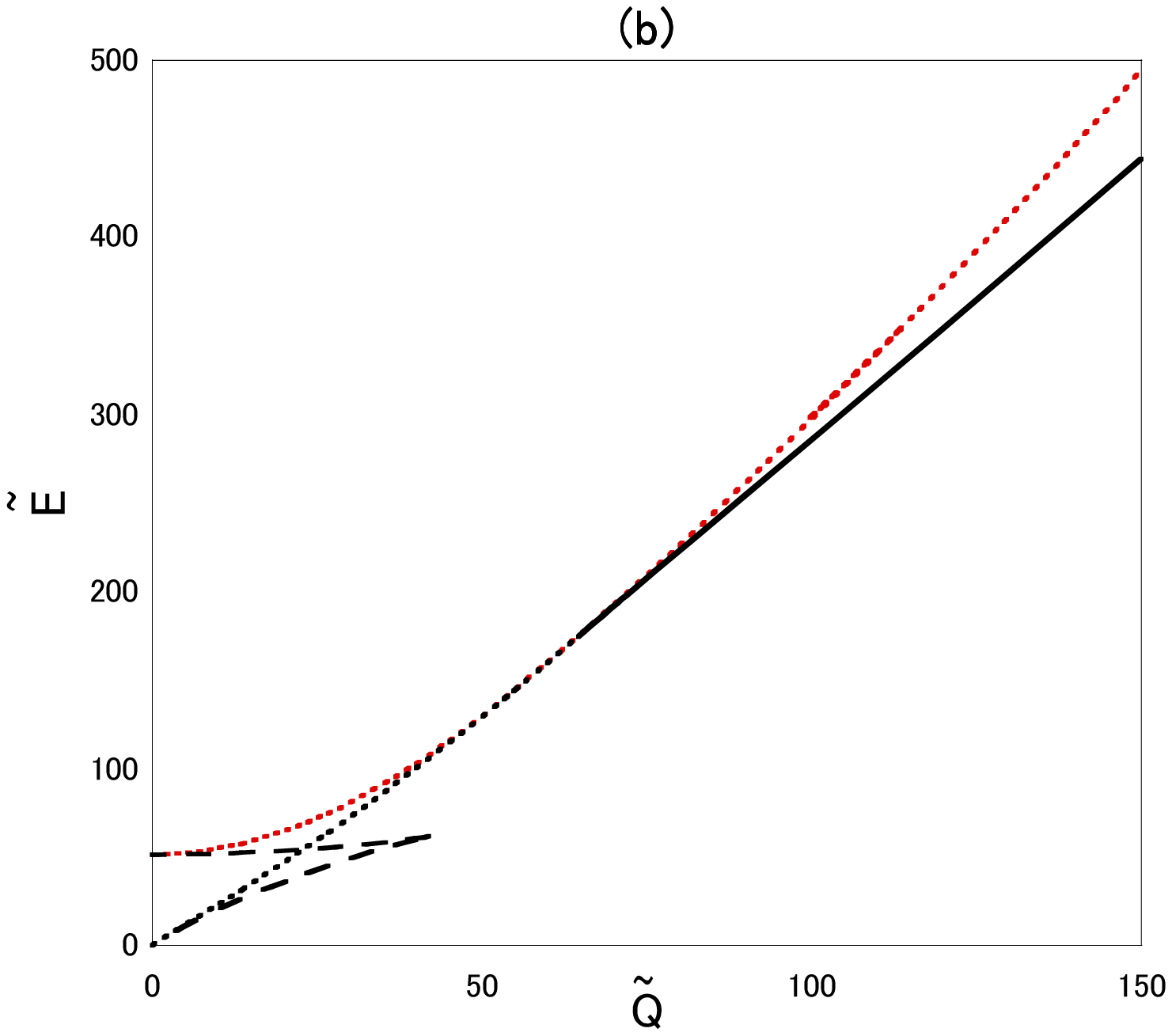,width=3.2in}
\caption{\label{K-106}
(a) $\tilde{\Omega}(0)$-$\tilde{\phi}(0)$ and (b) $\tilde{Q}$-$\tilde{E}$ relations for $K =-1.06$. 
The two sequences in red lines and in black lines are about to touch.
}
\end{figure}
\begin{figure}[htbp]
\psfig{file=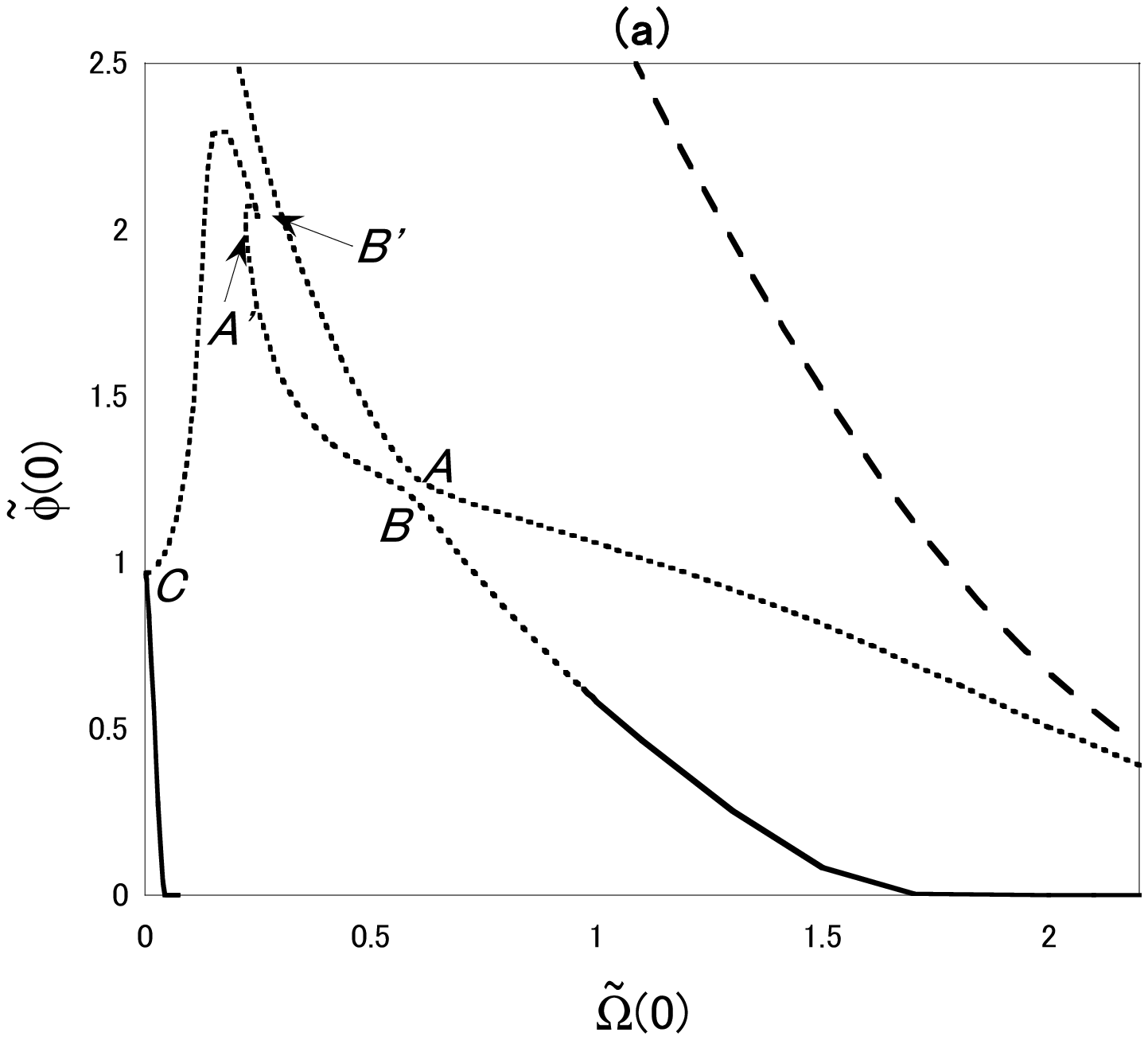,width=3.2in}
\psfig{file=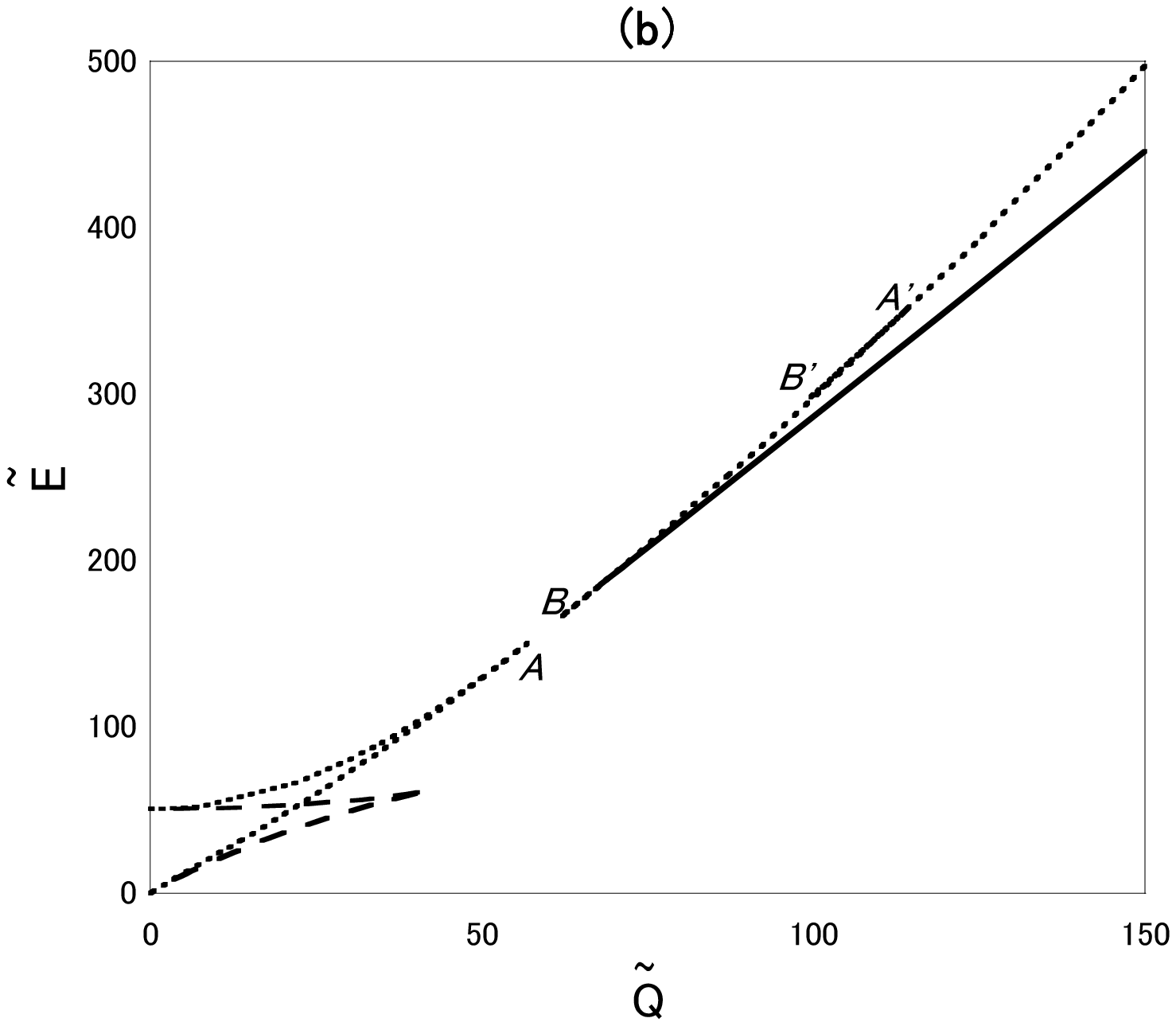,width=3.2in}
\caption{\label{K-107Omega-phi}
(a) $\tilde{\Omega}(0)$-$\tilde{\phi}(0)$ and (b) $\tilde{Q}$-$\tilde{E}$ relations with $K =-1.07$. 
The ``recombination" of the two sequences happens.
}
\end{figure}
\begin{figure}[htbp]
\psfig{file=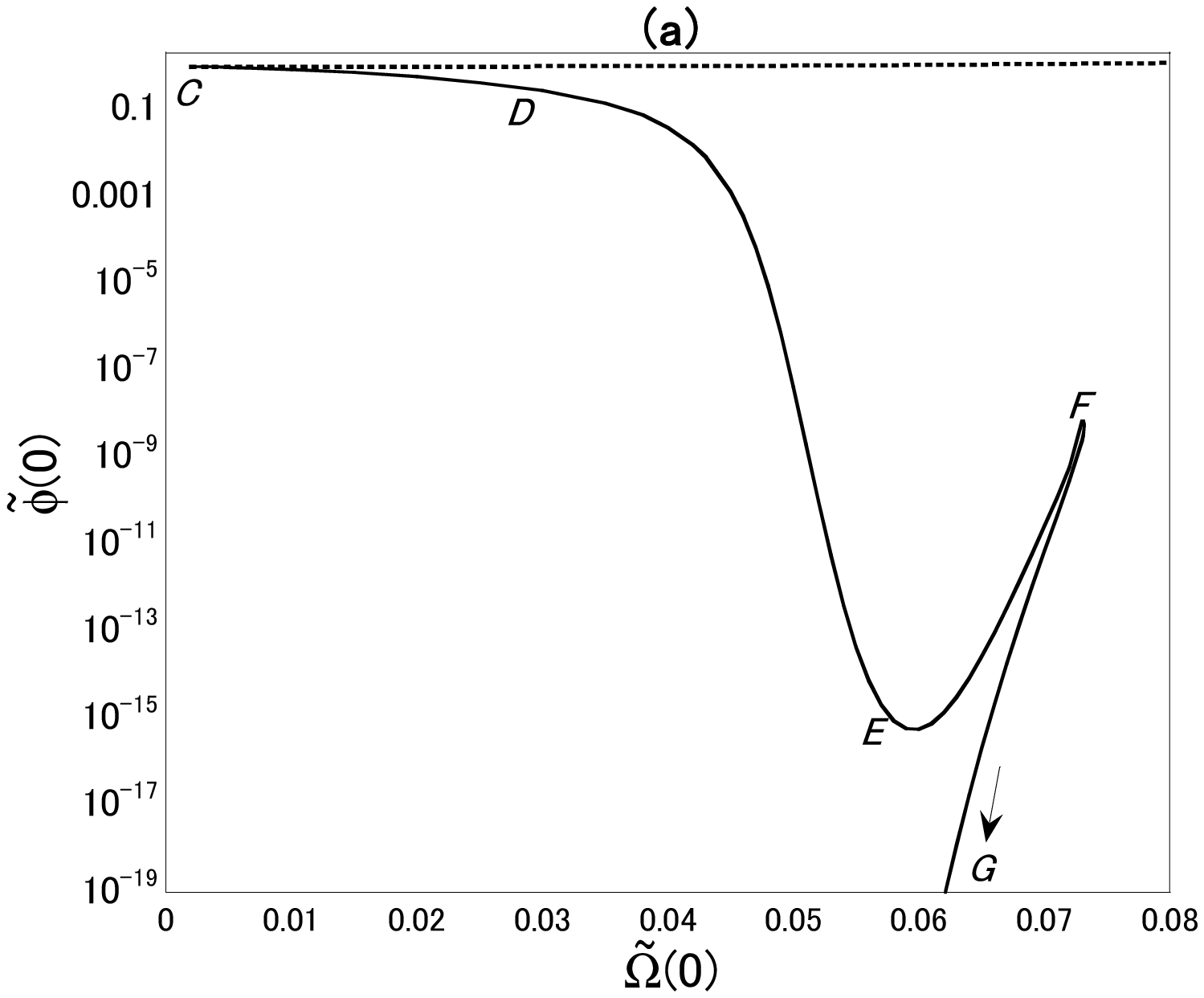,width=3.2in}
\psfig{file=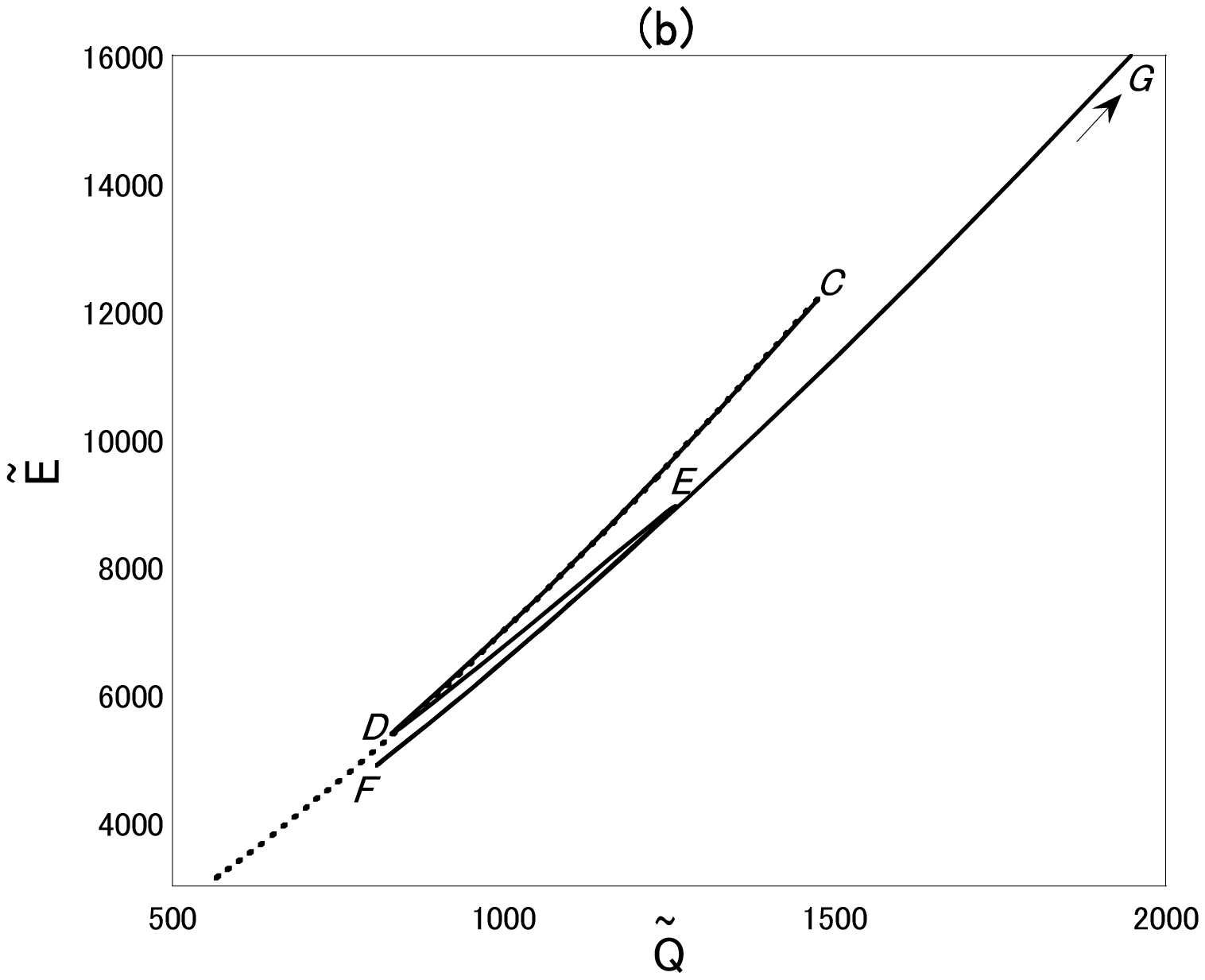,width=3.2in}
\caption{\label{k-107Q-E}
(a) $\tilde{\Omega}(0)$-$\tilde{\phi}(0)$ and (b) $\tilde{Q}$-$\tilde{E}$ relations for $K=-1.07$ and $\tilde{Q}>500$.
The dotted lines extend from Fig.\ \ref{K-107Omega-phi}.}
\end{figure}

Figures \ref{K-06} and \ref{K-04} show the $\tilde{\Omega}(0)$-$\tilde{\phi}(0)$ and $\tilde{Q}$-$\tilde{E}$ 
relations for $K =-0.6$ and $-0.4$, respectively. 
We find that, as $|K|$ decreases, the existing domain of the unstable solutions becomes small in the $\tilde\Omega(0)$-$\tilde\phi(0)$ plane and the two sequences leave away from each other.

A drastic change occurs between $K=-1.06$ and $K=-1.07$, as shown in Figs.\ \ref{K-106} and \ref{K-107Omega-phi}. 
As $|K|$ increases, the two sequences approach further; eventually at some point in $-1.07<K<-1.06$, the ``recombination" of the two sequences takes place. 
At first sight, this recombination looks strange; however, we can understand this phenomenon in a rational way as follows.
If we look at the existing domain of equilibrium solutions in the two-dimensional parameter space (say, the $\tilde{\Omega}(0)$-$\tilde{\phi}(0)$ or the $\tilde{Q}$-$\tilde{E}$ space) for fixed $K$, we see that there are two separate sequences of solutions.
However, if we consider the existing domain in the three-dimensional parameter by regarding $K$ as another parameter, it is described by a simply connected surface.
The ``recombination" of the two sequences is nothing but changing cross-sections of the same surface.

The solution sequence of the ordinary Q-balls, represented by the dashed line in Fig.\ \ref{K-107Omega-phi},
is analogous to the sequence including the point $A$.
The other sequence including the point $B$ has no counterpart of ordinary Q-balls.
The solutions in the branch $B$-$A'$-$B'$ are unstable, and there are two small cusps about $A'$-$B'$.
The  lower energy solutions in both sequences are stable; interestingly the two sequences of stable solutions are separated. There is no upper limit of $\tilde{Q}$.

\begin{figure}[htbp]
\psfig{file=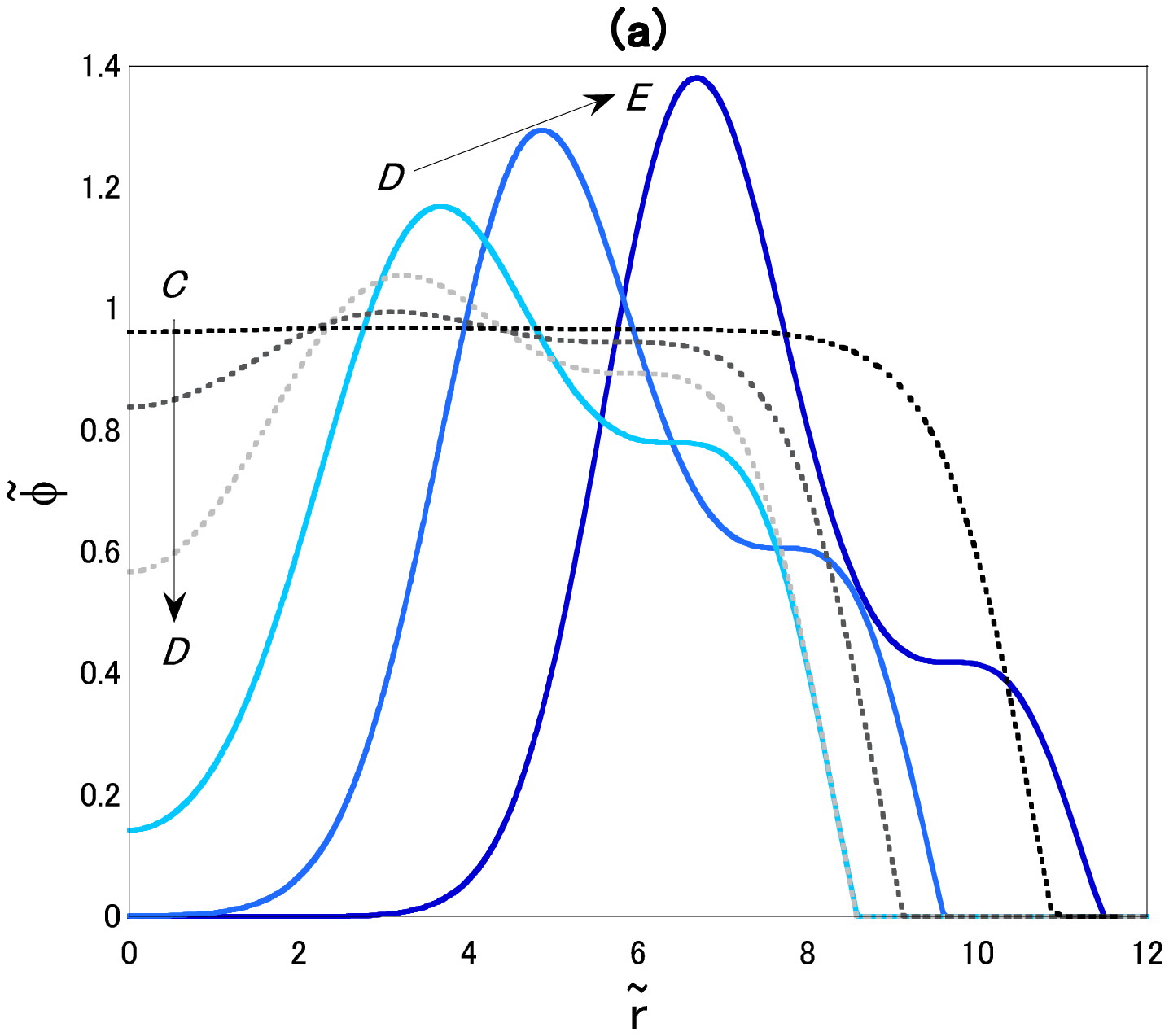,width=3.2in}
\psfig{file=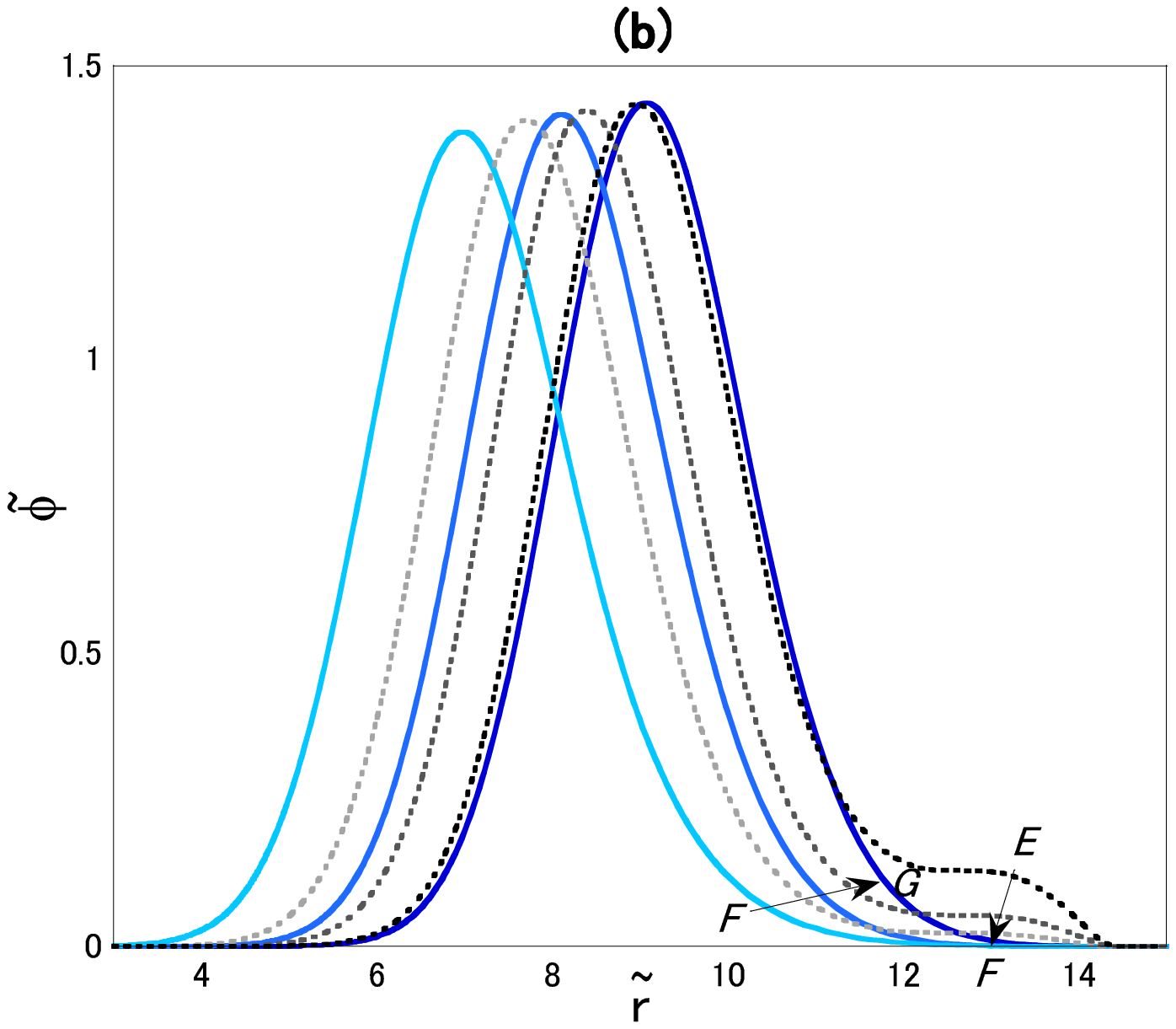,width=3.2in}
\caption{\label{K-107r-phi}
Field distributions of $\tilde{\phi}$ with $K=-1.07$ for (a)solutions $C$-$D$-$E$ and (b)solutions $E$-$F$-$G$. }
\end{figure}

As a common property for every $K$, there are sequences of cusp structures in the large $Q$ region for unstable solutions. 
We show (a) $\tilde{\Omega}(0)$-$\tilde{\phi}(0)$ and (b) $\tilde{Q}$-$\tilde{E}$ relations for very small 
$\tilde{\phi}(0)$ (and large $\tilde{Q}>500$) region for $K=-1.07$ in Fig.~\ref{k-107Q-E}.
Complicated structure appears along the sequence $C$ to $G$; there are several cusps about $C$-$D$-$E$-$F$. 
As shown in Fig~\ref{K-107r-phi}, field distributions in this region also have complicated structures. 
Beyond the point $F$, both $\tilde{\phi}_{\rm max}$ and $\tilde{r}_{\rm max}$ monotonically increase. 
It is interesting that small differences of boundary values $\tilde{\Omega}(0)$ and 
$\tilde{\phi}(0)$ result in such large differences in $\tilde{Q}$ and $\tilde{E}$.

\section{Summary and Discussions}

In many models of gauged Q-balls, which were studied in the literature, there are upper limits for charge and size of Q-balls due to repulsive Coulomb force.
As a cosmologically-motivated model which could allow for gauged Q-balls with large charge and size, we have considered the gravity-mediation-type model in the Affleck-Dine mechanism.
We have found that stable Q-balls with any amount of charge and size exist in this model as long as $K<0$.
As the electric charge $Q$ increases,  the field configuration of the scalar field becomes shell-like; 
because the charge is concentrated on the surface, the Coulomb force does not destroy the Q-ball configuration.
These properties are analogous to those in the V-shaped model, which was studied by Arod\'z and Lis \cite{Arodz}.
Because the V-shaped model is rather artificial, our results for the cosmologically-motivated model would be important if we consider gauged Q-balls as realistic dark matter model. 

We have also found that for each $K$ there is another sequence of unstable solutions, which is separated from the other sequence of the stable solutions.
As $|K|$ increases, the two sequences approach; eventually at some point in $-1.07<K<-1.06$, the ``recombination" of the two sequences takes place.
At first sight, this recombination looks strange.
However, if we consider the existing domain in the three-dimensional parameter by regarding $K$ as another parameter, it is described by a simply connected surface.
The ``recombination" of the two sequences is nothing but changing cross-sections of the same surface.

\acknowledgments
We would like to thank Kei-ichi Maeda for continuous encouragement. 
The numerical calculations were carried out on SX8 at  YITP in Kyoto University. 



\begin{thebibliography}{99}
\bibitem{Col85}S. Coleman, Nucl. Phys. {\bf B262}, 263 (1985). 
\bibitem{Kus97b-98}
A. Kusenko, Phys.Lett. B 405, 108 (1997) 108; Nucl. Phys. B (Proc. Suppl.) 62A-C, 248 (1998).
\bibitem{AD}I. Affleck and M. Dine, Nucl. Phys. B {\bf249} 361 (1985).
\bibitem{SUSY}
K. Enqvist and J. McDonald, Phys. Lett. B {\bf 425}, 309 (1998); Nucl. Phys. B {\bf 538}, 321 (1999);
S. Kasuya and M. Kawasaki, Phys. Rev. D {\bf62}, 023512 (2000).
\bibitem{SUSY-DM}
A. Kusenko and M. Shaposhnikov, Phys. Lett. B {\bf418}, 46 (1998); 
K. Enqvist, S. Kasuya and A. Mazumdar, Phys. Rev. D {\bf 66}, 043505 (2002); 
K. Enqvist, A. Mazumdar, Phys. Rept. {\bf 380}, 99 (2003); 
I. M. Shoemaker and A. Kusenko, Phys. Rev. D {\bf80}, 075021 (2009).
\bibitem{Kus98}A. Kusenko {\it et al.} Phys. Lett. B {\bf423} 104, (1998).
\bibitem{stability}
A. Kusenko, Phys. Lett. B {\bf404}, 285 (1997); {\bf406}, 26 (1997);
T. Multamaki and I. Vilja, Nucl. Phys. B {\bf 574}, 130 (2000);
M. Axenides, S. Komineas, L. Perivolaropoulos and M. Floratos, Phys. Rev. D {\bf 61}, 085006 (2000);
M. I. Tsumagari, E. J. Copeland, and P. M. Saffin, {\it ibid.}\ {\bf 78}, 065021 (2008).
\bibitem{PCS01}F. Paccetti Correia and M. G. Schmidt, Eur. Phys. J. {\bf C21}, 181 (2001).
\bibitem{SS}N. Sakai and M. Sasaki, Prog. of Theor. Phys., {\bf 119}, 929 (2008).
\bibitem{TS}T. Tamaki and N. Sakai, Phys. Rev. D {\bf81}, 124041 (2010); {\it ibid.}\ {\bf83}, 044027 (2011);
{\it ibid.}\ {\bf83}, 084046 (2011); {\it ibid.}\ {\bf84}, 044054 (2011).
\bibitem{SIN}N. Sakai, H. Ishihara and K. Nakao, Phys. Rev. D {\bf84}, 105022 (2011). 
\bibitem{TS2}T. Tamaki and N. Sakai, Phys. Rev. D {\bf86}, 105011 (2012). 
\bibitem{Lee}K. Lee, J. A. Stein-Schabes, R. Watkins, and L. W. Widrow, Phys. Rev. D {\bf39}, 1665 (1989).
\bibitem{AAFT}K. N. Anagnostopoulos, M. Axenides, E. G. Floratos, and N. Tetradis, Phys. Rev. D {\bf64}, 125006 (2001).
\bibitem{LHL}Xin-zhou Li, Jian-gang Hao, and Dao-jun Liu, J. Phys. A {\bf 34}, 1459 (2001).
\bibitem{DM}M. Deshaies-Jacques and R. MacKenzie, Can. J. Phys. {\bf 85}, 693 (2007).
\bibitem{Arodz}H. Arod\'z and J. Lis, Phys. Rev. D {\bf79}, 045002 (2009). 
\bibitem{PS78}For a review of catastrophe theory, see, e.g., 
T. Poston and I.N. Stewart, {\it Catastrophe Theory and Its Application}, Pitman (1978). 
\end{thebibliography}
\end{document}